\documentclass[floatfix,useAMS,usenatbib]{mn2e}
\usepackage{graphicx,epsfig}
\usepackage{multirow}
\usepackage{verbatim}
\usepackage[figuresright]{rotating}
\usepackage{txfonts}

\newcommand\araa{{ARA\&A}}
\newcommand\apj{{ApJ}}
\newcommand{\apjl}{ApJL}

\newcommand\aap{{A\&A}}
\newcommand\mnras{{MNRAS}}
\newcommand\aj{{AJ}}

\newcommand\nat{{Nature}}
\newcommand\nar{{New Astronomy Reviews}}

\title[The case of Tycho's supernova remnant]{Modeling the interaction of thermonuclear supernova remnants with circumstellar structures: The case of Tycho's supernova remnant }

\author[A. Chiotellis, D. Kosenko,  K.M. Schure, J. Vink and J.S. Kaastra ]{A. Chiotellis$^{1}$\thanks{E-mail:
A.Chiotellis@uva.nl}, D. Kosenko$^{2,3}$, K.M. Schure$^4$, J. Vink$^{1}$ and J.S. Kaastra$^5$\\ 
$^{1}$Astronomical Institute ``Anton Pannekoek", University of Amsterdam, P.O. Box 94249, 1090 GE Amsterdam, The Netherlands\\
$^{2}$Service d'Astrophysique,  L'Orme des Merisiers, CEA-Saclay, F-91191, Gif sur Yvette, Cedex, France\\
$^{3}$Sternberg Astronomical Institute (MSU), 119992, Universitetsky pr.13., Moscow, Russia \\
$^{4}$Department of Physics, University of Oxford,  Clarendon Laboratory, Parks Road, Oxford OX1 3PU, United Kingdom\\
$^{5}$SRON Netherlands Institute for Space Research, Sorbonnelaan 2, 3584 CA Utrecht, The Netherlands}

\begin{document}
\date{\ldots;\ldots}
\pagerange{\pageref{firstpage}--\pageref{lastpage}}\pubyear{2013}
\maketitle
\label{firstpage}

\begin{abstract}
The well-established Type Ia remnant of Tycho's supernova (SN 1572) reveals discrepant ambient medium density estimates based on either the measured dynamics
or on the X-ray emission properties. This discrepancy can potentially be solved by assuming that the supernova remnant (SNR) shock
initially moved through a stellar wind bubble, but is currently evolving in the uniform interstellar medium with a relatively
low density.

We investigate this scenario by combining hydrodynamical simulations of the wind-loss phase and the supernova remnant evolution
with a coupled X-ray emission model, which includes non-equilibrium ionization. For the explosion models
we use the well-known W7 deflagration model and the delayed detonation model that was previously shown to provide good fits
to the X-ray emission of Tycho's SNR.

Our simulations confirm that a uniform ambient density cannot simultaneously reproduce the dynamical and X-ray emission properties
of Tycho. In contrast,  models that considered that the remnant was evolving in a dense, but small, wind bubble reproduce reasonably well both the measured
X-ray emission spectrum and the expansion parameter of Tycho's SNR. Finally, we discuss possible mass loss scenarios in the context of single- and double-degenerate models which possible could form such a small dense wind bubble.
\end{abstract}


\section{Introduction}\label{Sec:Intro}

Type Ia supernovae (SNe Ia) are believed to be the result of the thermonuclear explosion of a carbon oxygen white dwarf (CO WD) which accretes mass from a binary companion  \citep{WI73}. The mass transfer could be  through Roche lobe overflow, through the stellar wind of a non-degenerate donor star  (single degenerate -- SD -- scenario), or through the merger with an other WD (double degenerate -- DD -- scenario). 
Despite decades of research, there are still major questions about
the nature of the explosion mechanism, the mass transfer process and  the nature of the donor star 
\citep[see the reviews by][]{WangHan12,hillebrandt00, Livio2000}.
 Regarding the nature of the progenitor system, from the observational side the SD scenario has problems in explaining the lack of radio and X-ray emission from the Type Ia blast wave interacting with the wind of the donor star \citep[e.g][]{chomiuk12,Immler2006}, the lack of super soft X-ray radiation from elliptical galaxies expected by accreting white dwarfs \citep{gilfanov10}, and the lack of surviving donor stars in Type Ia 
supernova remnants \citep[e.g.][]{schaefer12}. On the other hand, some Type Ia SNe show evidence for local blue-shifted sodium absorption,
which is best explained by the presence of shells caused by outflows from a SD progenitor system \citep[e.g.][]{dilday2012,sternberg11,Patat07}.

A useful method to clarify the nature of SNe Ia is the study of their supernova remnants (SNRs). The progenitors of Type Ia supernovae  can modify the ambient medium  either through mass outflows during the binary evolution,  or through the ionizing radiation that can accompany accretion.  The subsequent interaction of the supernova forward shock with  the modified circumstellar medium will affect the evolution of the resulting SNR. In this sense, traces of the nature and evolution of SN Ia progenitors  will be reflected in the observed morphological, dynamical and emission properties of the nearby  Ia SNRs. 
Although, in general, the remnants of SNe Ia tend to be more spherical than their core collapse counterparts  \citep{lopez09}, something that indicates that the surrounding of the first is not substantially modified by their progenitors, several observations of nearby Ia SNRs show direct or indirect evidence of a SN ejecta - circumstellar medium (CSM) interaction. 
The most famous examples are the remnants of the historical SNe Ia of Kepler (SN 1604) and RCW 86.   It has been shown that the observed characteristics of these remnants  can be well reproduced considering a  SD progenitor scenario where, for the case of Kepler's SNR, the SN explosion occurred in a dense, bow-shaped bubble formed by the wind of the AGB  donor star 
\citep{chiotellis12,patnaude12,Burkey2013}, 
while for the case of RCW 86, which has been argued to be a Type Ia SNR \citep{williams11}, there is evidence that
it evolves in a low-density cavity carved by the mass outflows emanating from  the progenitor WD 
\citep{williams11,badenes07}.  
Other examples are the mature Ia SNRs DEM L238, and DEM L249 in the LMC which have enhanced, Fe-rich,  central emission, indicating
high central densities, which
can possibly be explained by an early interaction of the SNR with a dense CSM \citep{borkowski06}.

The historical SNR of Tycho (SN 1572) is an other Galactic SNR which has been classified as a Type Ia, 
based initially on its ejecta chemical composition \citep{hwang97}, but the Type Ia nature  has now been unambiguously confirmed based
on the light echo spectrum \citep{krause08b}. The SNRs angular radius is $\sim$250\arcsec, but its distance is uncertain, with estimates ranging 
between $\sim 1.8 - 5 $ kpc \citep[see Fig. 6 of][]{hayato2010}.    In contrast  with the Kepler and RCW 86, Tycho's SNR (hereafter Tycho) is characterized by an almost quasi-circular shape with smooth rims, indicating that the SNR is currently evolving in a more or less homogeneous ambient medium (AM).

\begin{figure}
	\centering
		\includegraphics[trim=10 10 10 10,clip=true,width=\columnwidth]{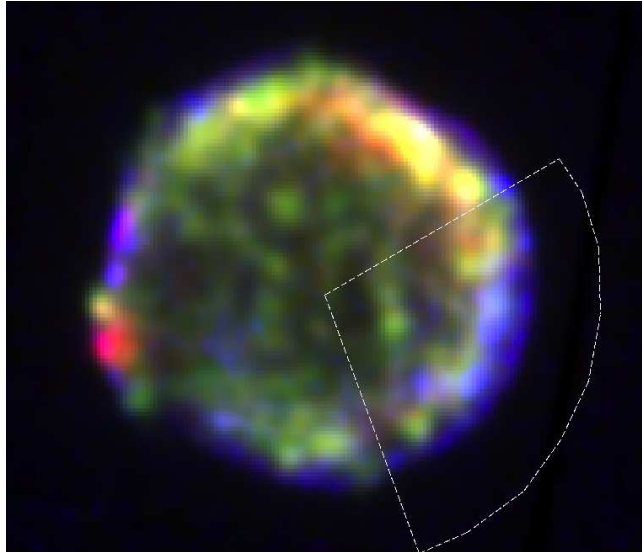}
	\caption{XMM Newton EPIC merged MOS1 and MOS2 image of Tycho SNR (OBS ID~0096210101).  The angular sector of $200^o-300^o$ is outlined by the white, dashed line. The red, green, and blue colors (online version) correspond to FeL emission (0.9-1.1 kev),
Si XIII emission (1.7-2.0 keV) and continuum emission (4.2-6 keV), respectively.
}
	\label{fig:mos-sky}
\end{figure}

Under the assumption that Tycho is expanding in a homogeneous medium, \citet{badenes06} find a good match between the synthetic X-ray spectra from their model and the observed integrated spectra of the southwest-west (SWW) region of the remnant ($\rm 200^o - 300^o$ counterclockwise starting from the north; see Fig. \ref{fig:mos-sky}),  for an interstellar density of  $\sim 0.9~\rm cm^{-3}$.  However, the lack of  thermal X-ray emission from the shocked ambient medium places an upper limit  on the ambient medium density of   $\sim 0.3~ \rm cm^{-3}$ \citep[with an uncertainty factor of 2,][]{cassam07} in agreement with mid-infrared measurements of Tycho which reveal an averaged AM density of  $\sim 0.1 - 0.2 ~\rm cm^{-3}$ \citep{williams13}. Finally, an ambient density of  $\sim 0.1~ \rm cm^{-3}$ is required in order to match the expansion parameter  $m = V_s/(R_s/t_{SNR})$ from SNR evolution models \citep{dwarkadas00} with the expansion parameter derived from X-ray \citep{katsuda10} and radio \citep{reynoso97} observations of the shock radius ($R_s$) and velocity ($V_s$) at the SWW region of the remnant.

 The physical interpretation of this discrepancy between various ambient medium densities for Tycho
is the following:  On the one hand the high velocity of Tycho's forward shock ($\sim 0.35~ \rm arcsec~yr^{-1}$)  and the lack of thermal emission at the shocked ambient medium shell demand a low density environment in which the remnant is evolving. On the other hand, Tycho's X-ray spectrum is characterized by a high ionization age ($\tau= \int n_e dt$, with $n_e$ the electron density of the shocked ejecta), requiring high shocked-ejecta densities. This in  turn requires a  high ambient medium density that surrounds the remnant. Thus,  models that consider a constant density ambient medium around Tycho  have difficulties explaining simultaneously the 
observed ionization age with the measured dynamics and the thermal X-ray flux from the SNR.
%emission that reveal the X-rays observations of the SNR and vice versa. 
 Based on these considerations  \citet{dwarkadas98} and \citet{katsuda10} argued that a more complex circumstellar structure could possibly
reconcile the different estimates of the ambient medium densities.

Here we study this hypothesis on the premise that Tycho's SNR has interacted with a circumstellar wind bubble, formed by a period of continuous mass outflow from the progenitor binary, the characteristics of which we will determine in order to create a match between dynamics and the expected X-ray emission. We use different codes in order to get the best results for the dynamics (AMRVAC),  and X-ray spectra (SUPREMA and SPEX).

\section{The impact of  the ambient medium on the SNR evolution }\label{sec:study}

In this section we study such a SN ejecta -- wind bubble interaction  in light of the dynamical and emission properties of the resulting SNR. We compare the properties of such a SNR with  those of a SNR that evolves in a homogeneous medium from the start  in a qualitative sense, before we explore the potential applicability of such a SNR-wind bubble model for Tycho in the next Section.  

We employ the hydrodynamical code %of the 
AMRVAC % framework 
\citep{Keppens03}, and perform the simulation on  a one dimensional (1D) grid assuming spherical symmetry. The method that we follow includes two steps. First, we simulate the wind bubble taking the wind of the progenitor as a constant inflow at the inner boundary of the grid with a density profile of $\rho= \dot{M}/(4 \pi u_w r^2)$ and a momentum per unit volume of  $m_r= \rho \times u_w$, where $\dot{M}$ is the mass loss rate of the wind, $u_w$ the wind terminal velocity and $r$ the radial distance from the mass losing object.   In the second stage, we introduce the supernova ejecta in the center of the wind bubble. The code has the advantage of the adaptive mesh strategy,  which means that it is able to refine the grid locally where large gradients in density and/or momentum take place. As a result, the discontinuities that characterize the SNR (forward shock -- FS , reverse shock -- RS and contact discontinuity -- CD) are well tracked and we extract their dynamical properties  (radius, velocity, expansion parameter) during the whole evolution of the remnant. Finally,  radiative cooling, which may be important for the evolution of the stellar wind bubble,
 is included using the cooling curve introduced by \citet{Schure09}.

\subsection{The evolution and collision of a SNR with a wind bubble}\label{sec: SNR_wind}

\begin{figure}
	\centering
		\includegraphics[trim=17 17 17 17,clip=true,width=\columnwidth]{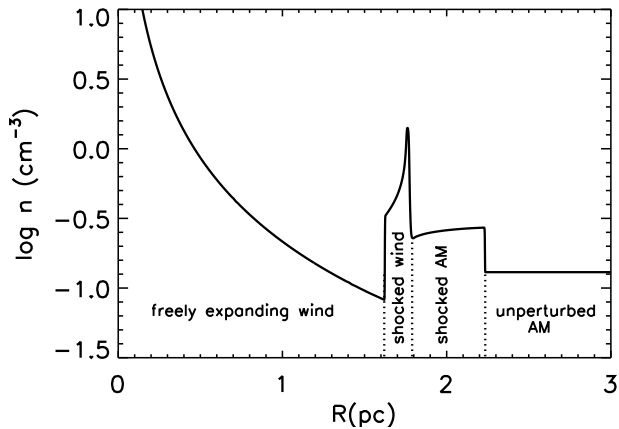}
	\caption{The density profile of the  four zone structure of the wind bubble. The wind properties of the specific simulation are:  mass loss rate $\dot{M}= 10^{-6}~ \rm{M_{\odot}~yr^{-1}}$ , wind terminal velocity $u_w= 10 ~\rm{km~s^{-1}}$ and interstellar medium density $n_{ism} = 0.13~ \rm{cm^{-3}}$. The age of the wind bubble is 0.25 Myr and the temperature of the wind and ISM is  $T= 10^3$ K.  Note that these values are typical
for our model for Tycho, but for illustrative purposes we made 
some modifications to allow for a better distinction
between the freely expanding wind region, the shocked wind region, and the shocked ambient medium shell.
  }
	\label{fig:wind_prof}
\end{figure}

 We illustrate the general effects of the progenitor system on the circumstellar medium
in Figure \ref{fig:wind_prof}, 
which shows the density profile of a stellar wind simulation. 
The wind parameters that have been used in this example are: mass loss rate $\dot{M}= 10^{-6}~ \rm{M_{\odot}~yr^{-1}}$, wind terminal velocity $u_w= 10 ~\rm{km~s^{-1}}$ and interstellar medium density $n_{ism} = 0.13 ~\rm{cm^{-3}}$.  At this specific snapshot the age of the wind bubble is 0.25~Myr. Due to the supersonic expansion of the wind bubble and the interaction of it with the ambient medium,  two shock waves are formed (similarly to the SNR evolution): the forward shock  and the termination shock that move outwards and inward  in the freely expanding wind rest frame, respectively. The resulting  structure of the wind bubble is characterized by four regions (starting from inwards): a) the freely expanding wind, where the density ($\rho$), scales with the distance from the mass  losing star ($r$) as $\rho \propto r^{-2} $,  b) the shocked wind material swept up by the termination shock, c) the shell of the shocked ambient medium, which contains the material being compressed by the wind's forward shock d) the undisturbed ambient medium.

\begin{figure*}%[htbp]       & trim l b r t
\begin{center}$
\begin{array}{cc}
\includegraphics[trim=45 10 18 10,clip=false,width=70mm,angle=0]{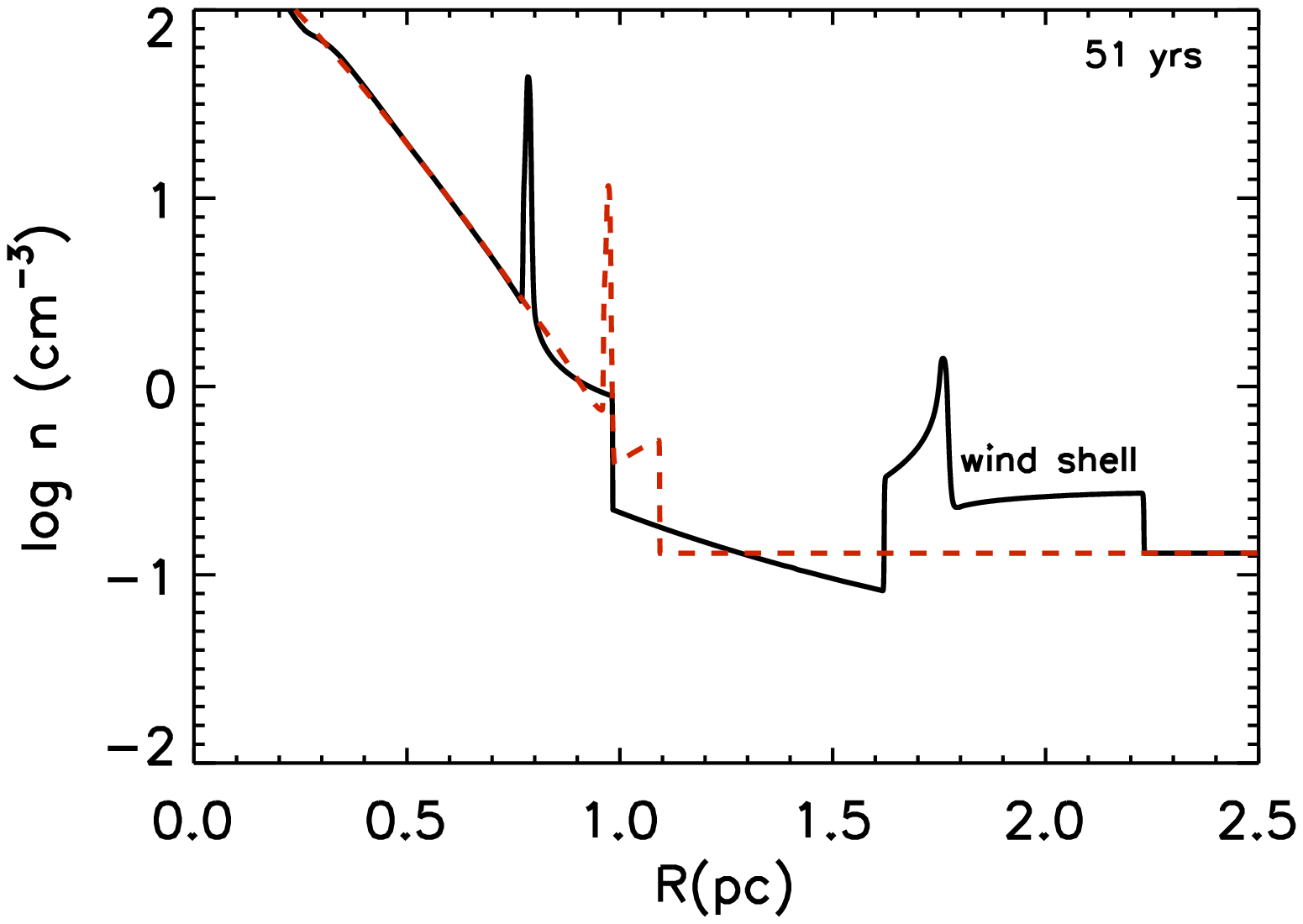} &
\includegraphics[trim=45 10 18 10,clip=false,width=70mm,angle=0]{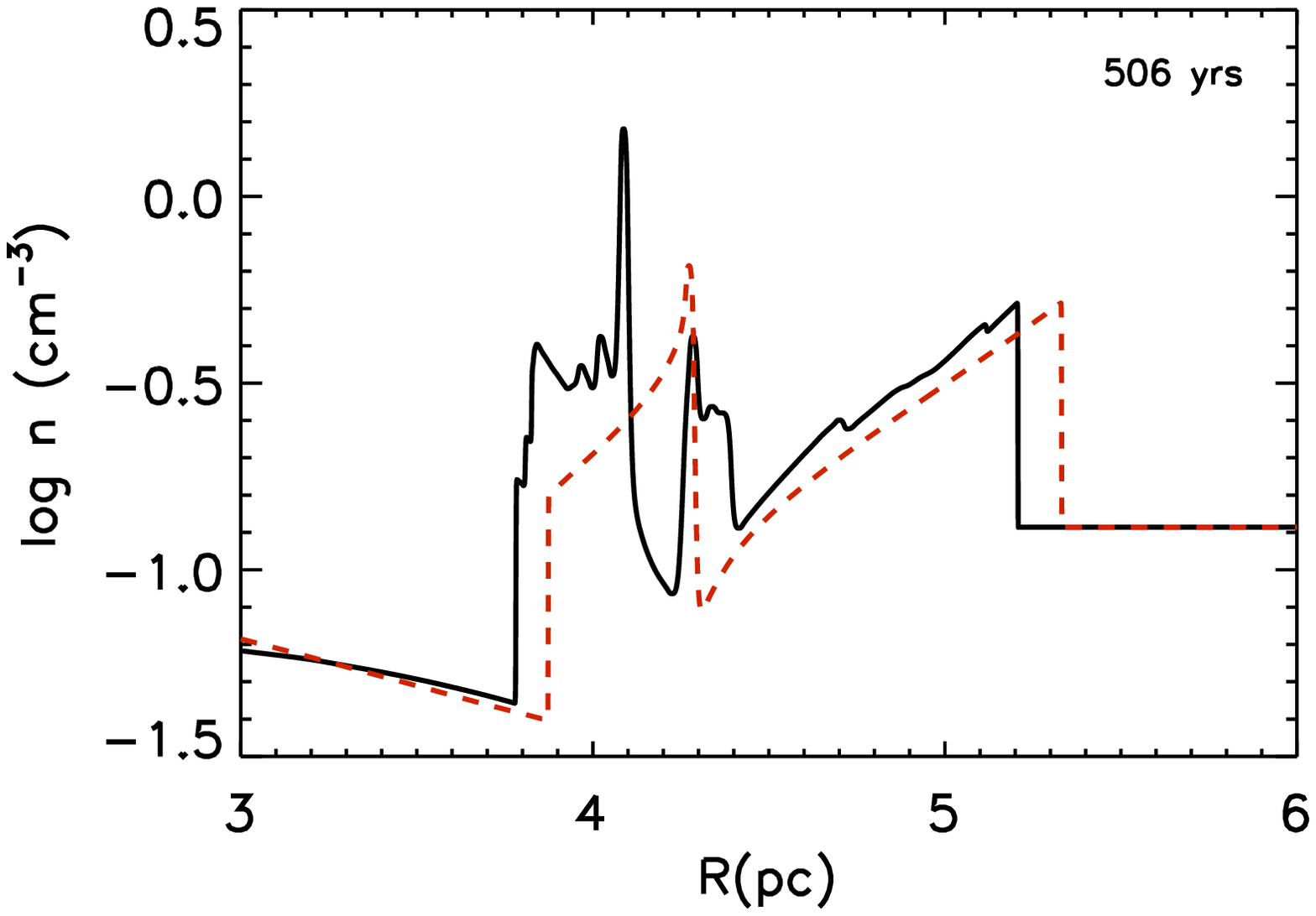} 
\\
\includegraphics[trim=45 10 18 10,clip=false,width=70mm,angle=0]{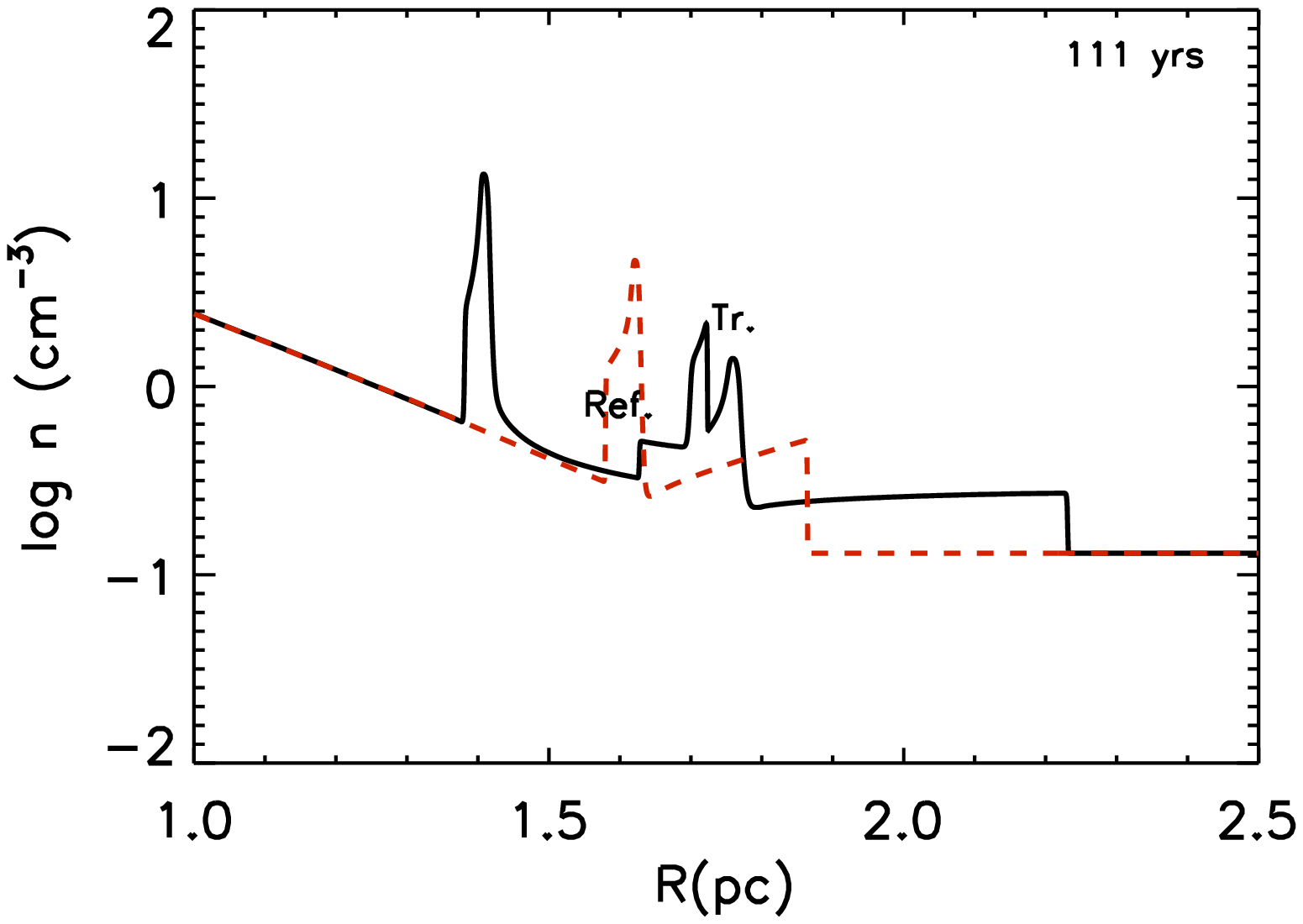} &
\includegraphics[trim=45 10 18 10,clip=false,width=70mm,angle=0]{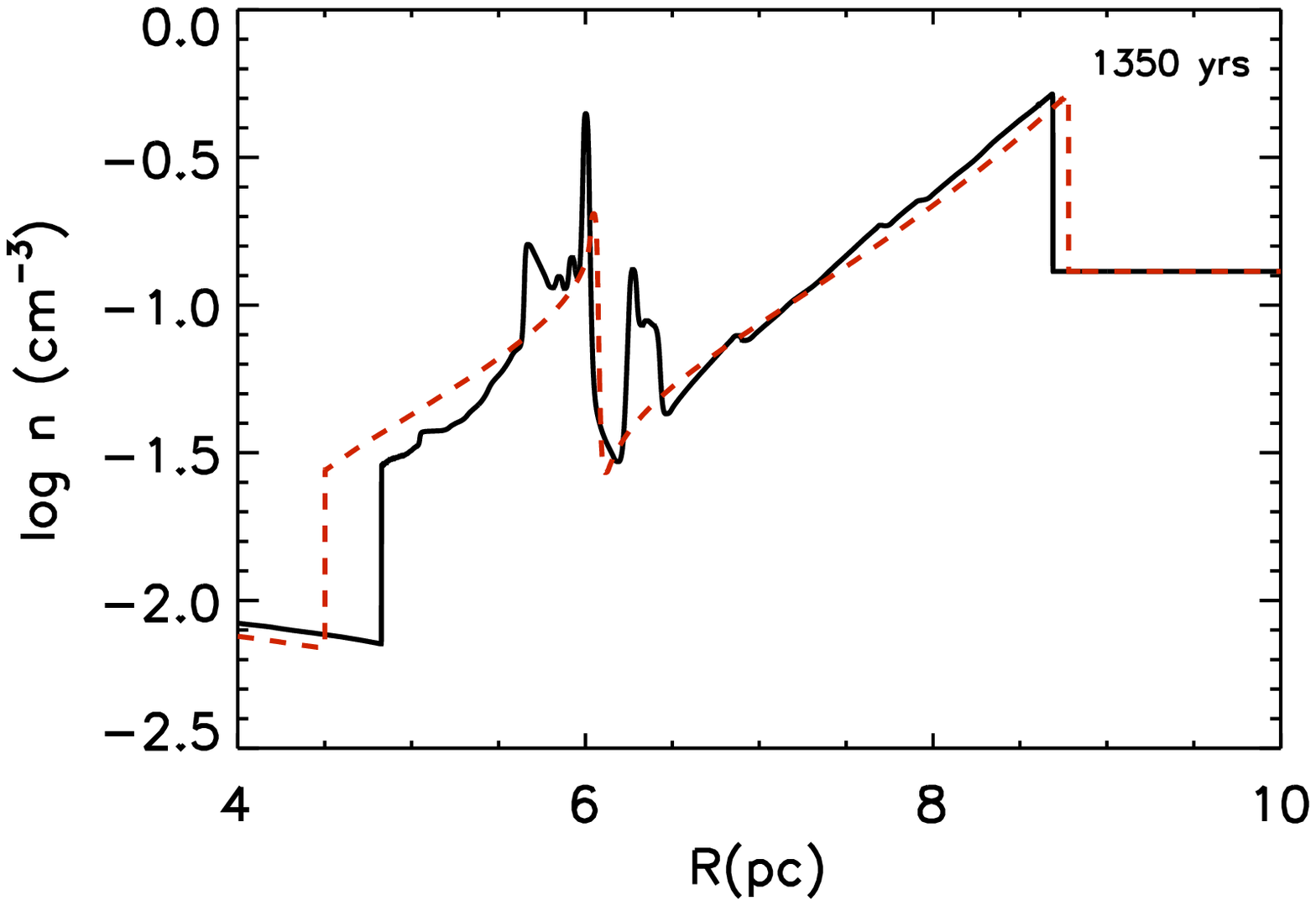}
\\ 
\includegraphics[trim=45 10 18 10,clip=false,width=70mm,angle=0]{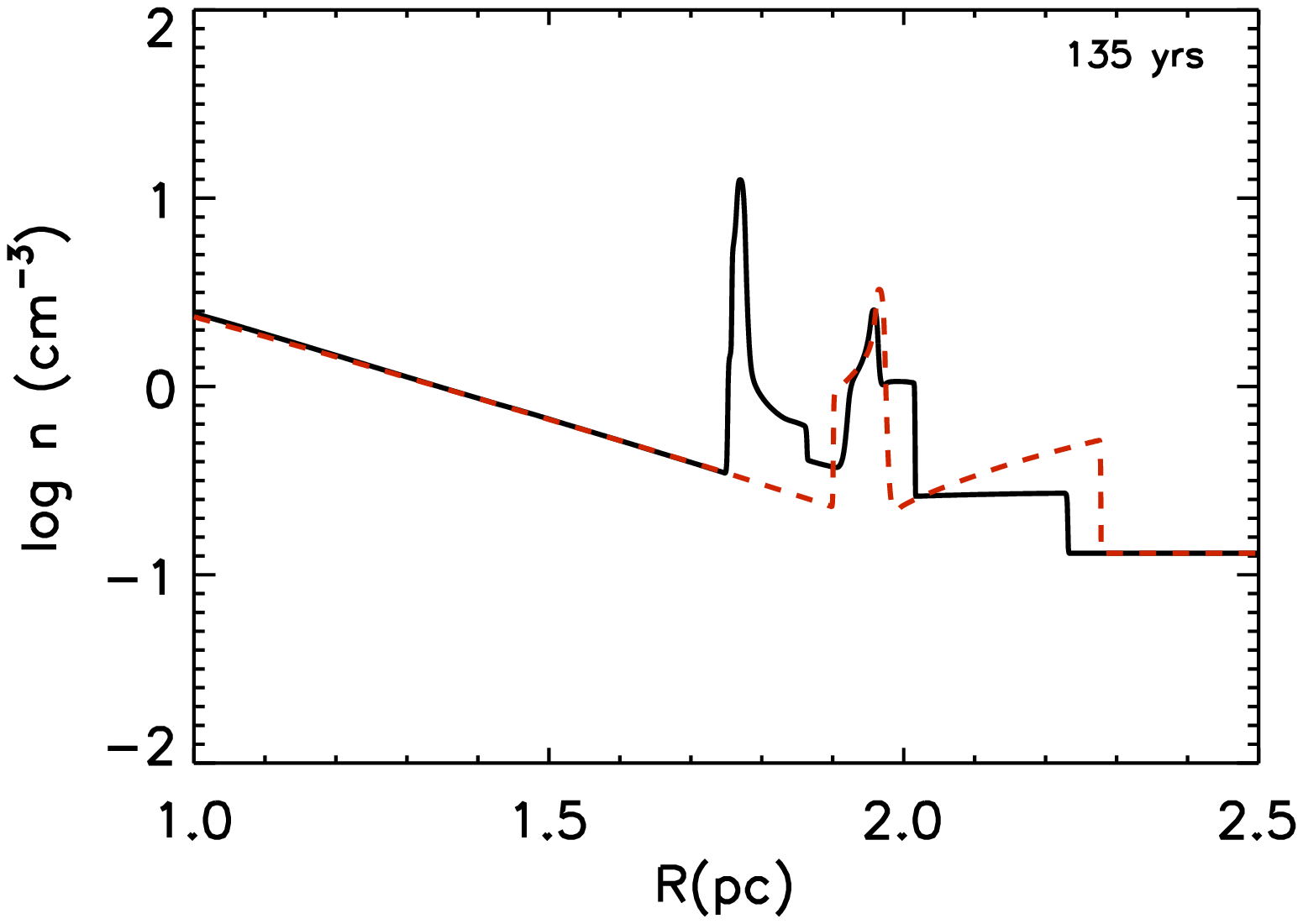} &
\includegraphics[trim=45 10 18 10,clip=false,width=70mm,angle=0]{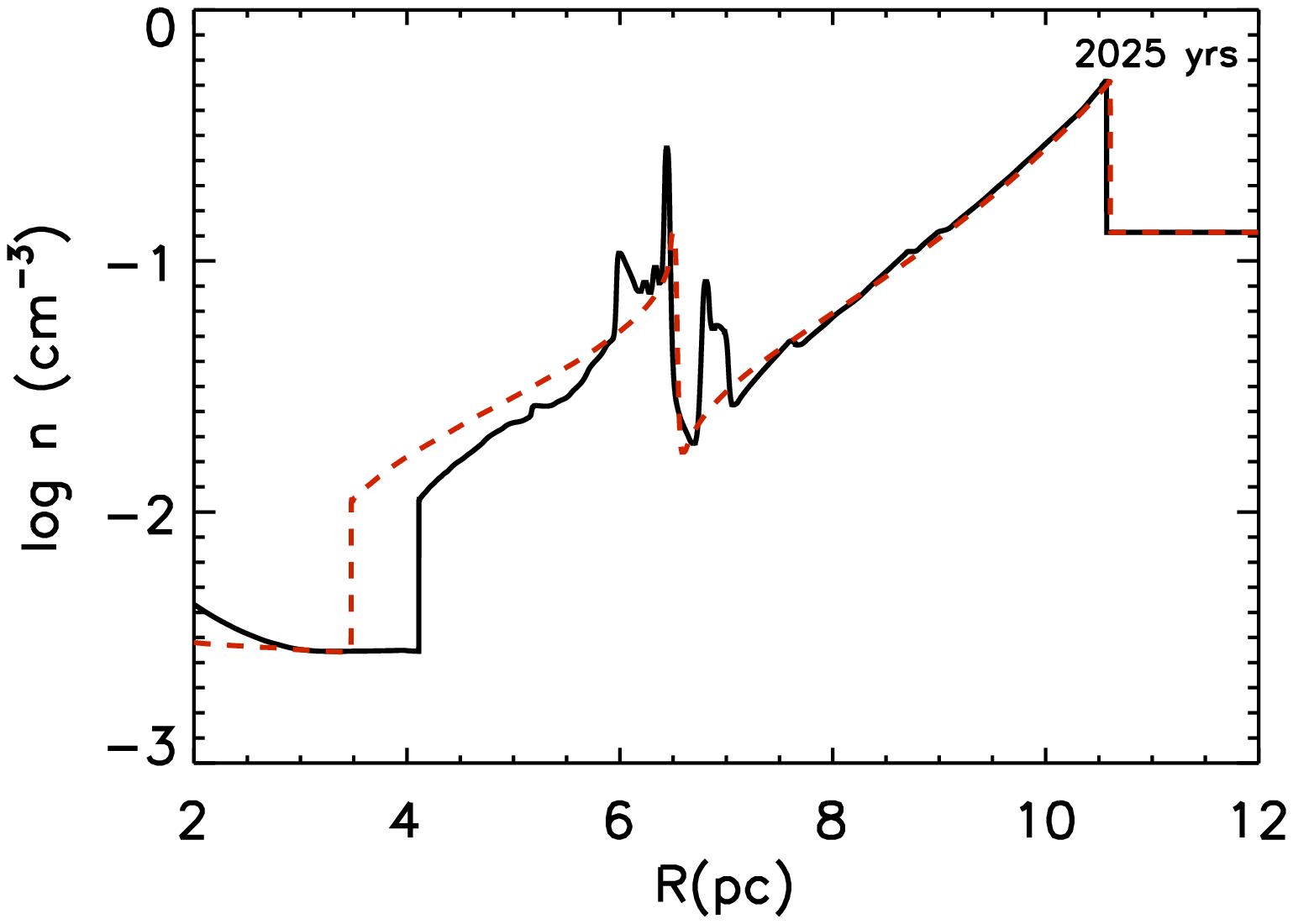}

\end{array}$
\end{center}
\caption{Comparison of the density profile of a SNR  evolving in the wind bubble of Fig. \ref{fig:wind_prof} (black solid line) with this of a SNR evolving in a homogeneous medium from the start (red dashed line) for several snapshots. The labels `Tr.' and `Ref.' refer to the transmitted and reflected shock respectively.    }
  \label{fig:SNR_comp}
\end{figure*}

\begin{figure*}%[htbp]       & trim l b r t
\begin{center}$
\begin{array}{ccc}
\includegraphics[trim=45 10 10 10,clip=true,width=55mm,angle=0]{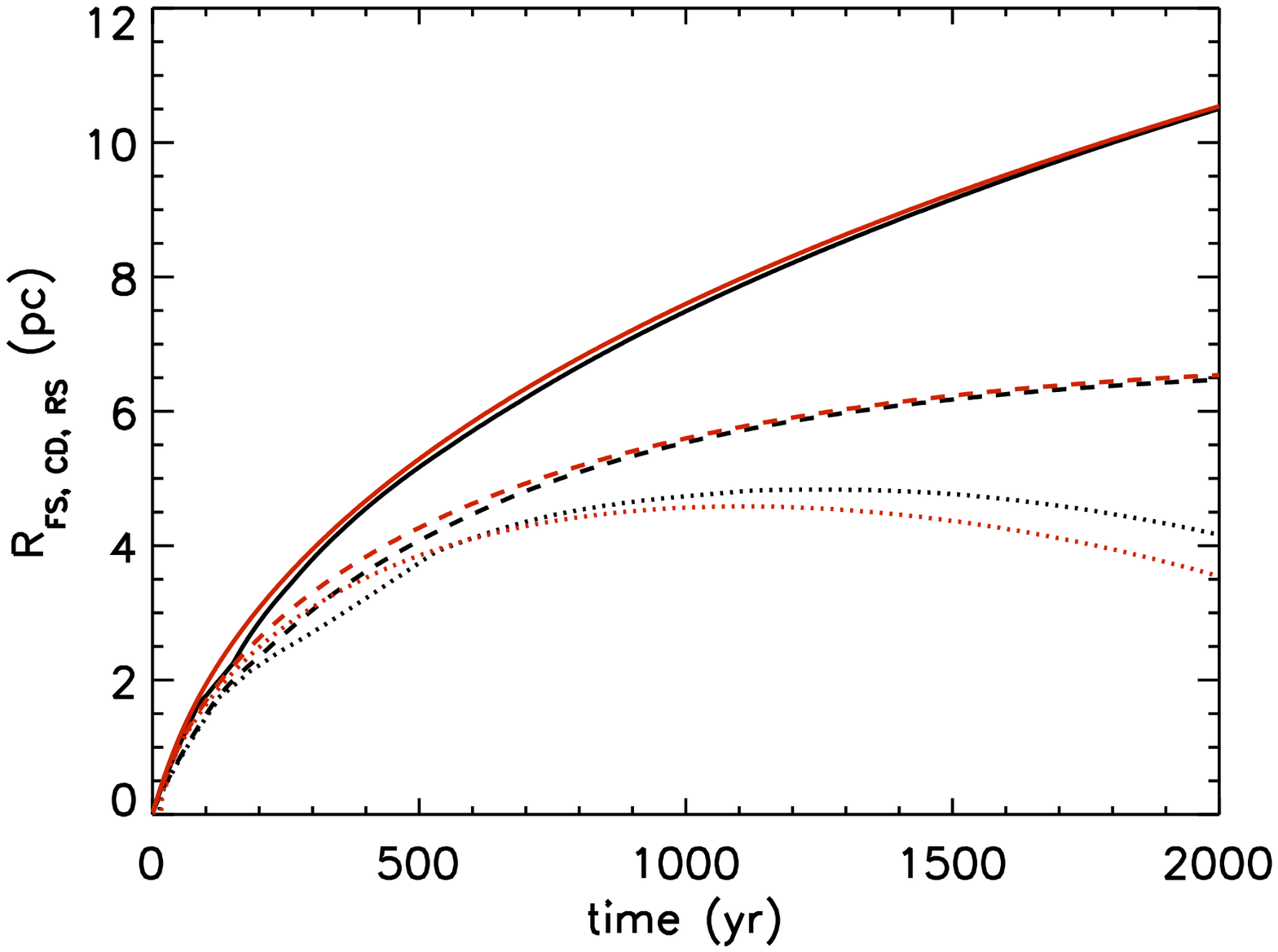} &
\includegraphics[trim=45 10 10 10,clip=true,width=55mm,angle=0]{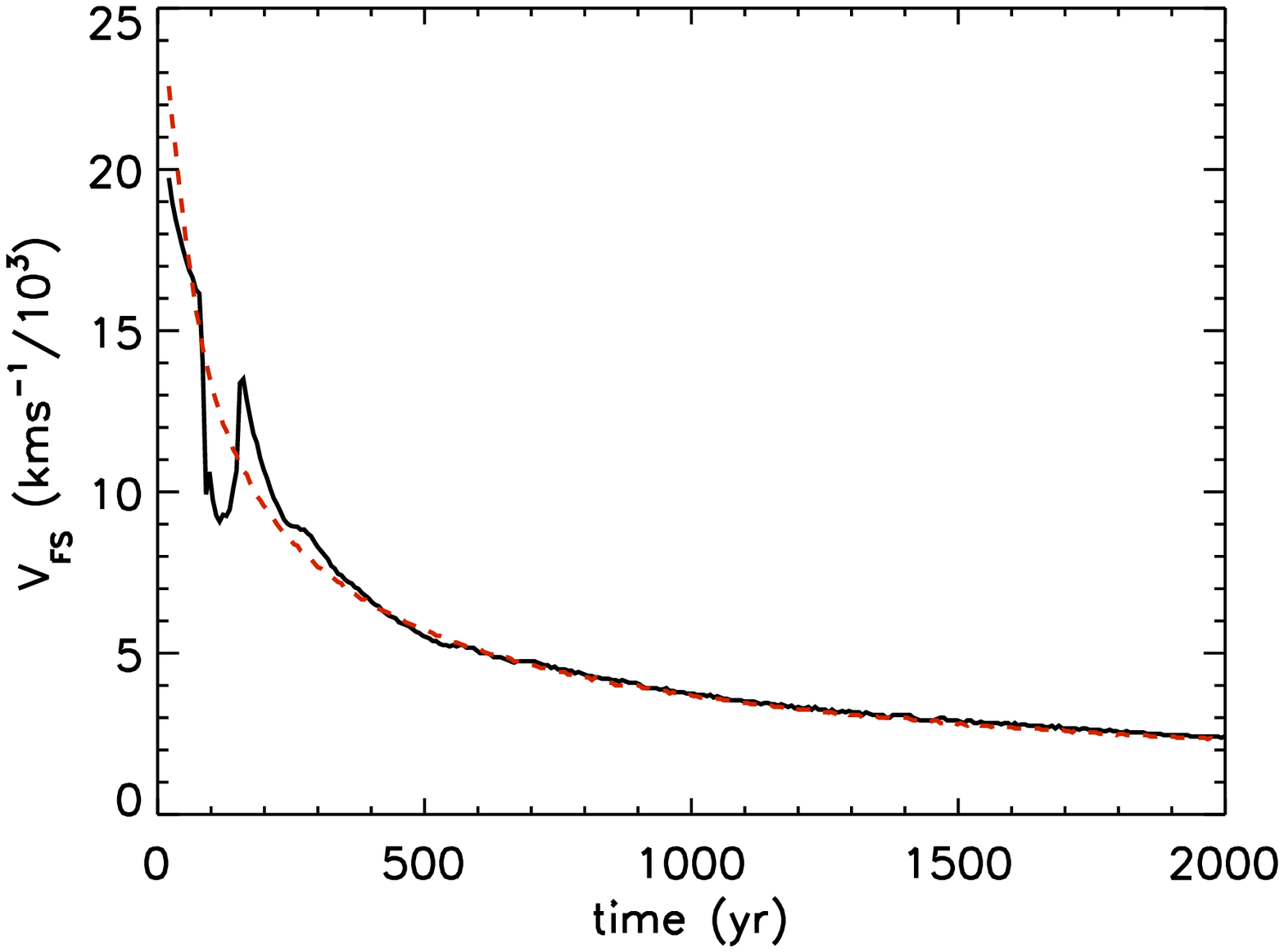} &
\includegraphics[trim=45 10 10 10,clip=true,width=55mm,angle=0]{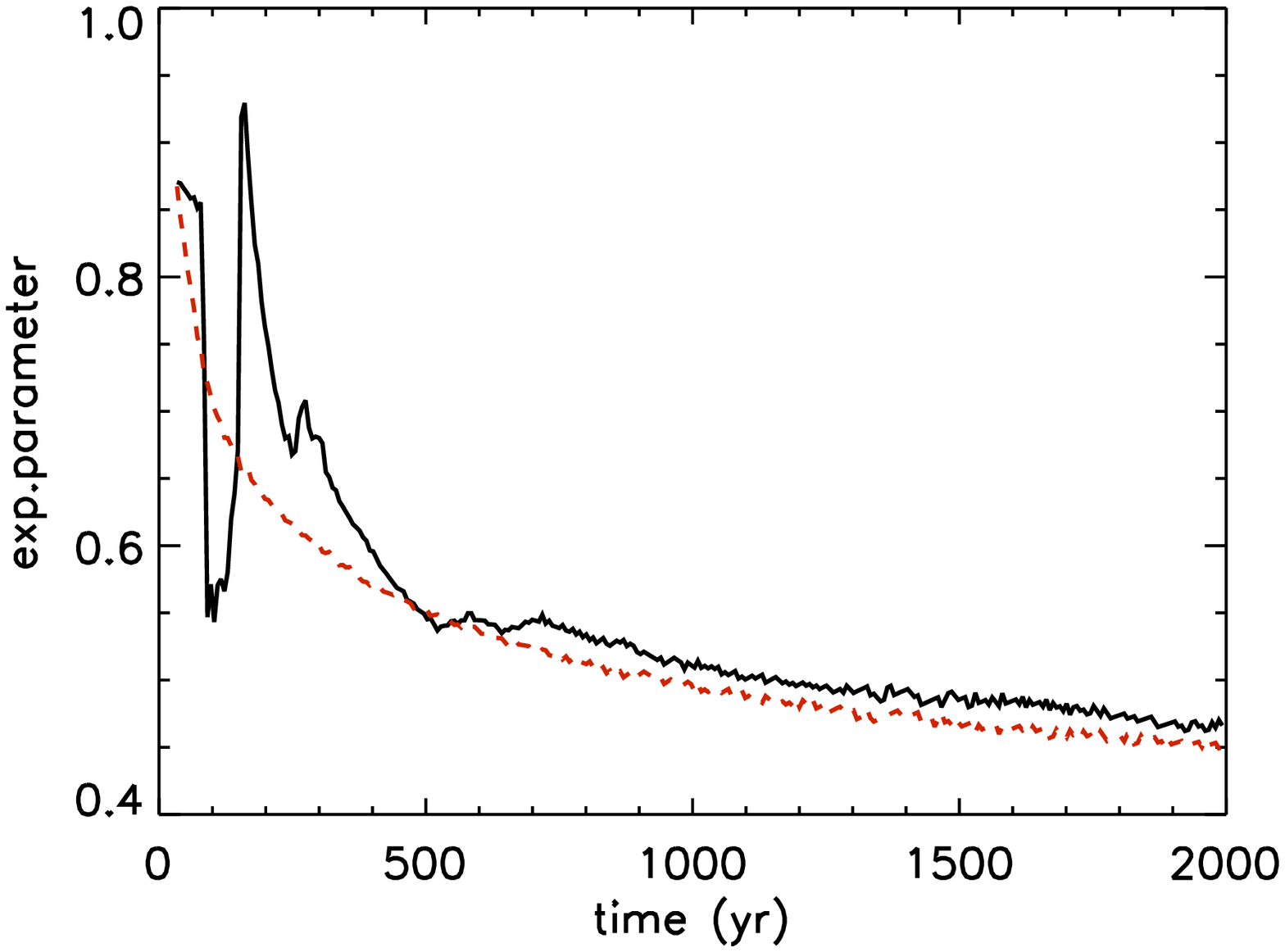} \\
\end{array}$
\end{center}
\caption{Dynamical properties of the  SNR with the wind bubble interaction history (black lines). Comparison with the ISM case (red lines).  Left: The radius of the forward shock (solid line), contact discontinuity (dashed line) and reverse shock (dotted line) as a function of time. Middle: the velocity  of the forward shock.  Right: the expansion parameter  of the forward shock as function of time.  }
 \label{fig:SNR_dynamics_comp}
\end{figure*}

It is into this specific ambient medium  that we introduce the supernova ejecta,  
with an energy of $1.3 \times 10^{51}$~erg, typical for SNe Ia \citep[][]{badenes05},
 and a mass  of $1.38~ \rm{M_{\odot}}$, following an exponential density profile. The outer edge of the ejecta is at $\sim 3 \times 10^3~$AU, corresponding to a starting time for
the calculations of $\sim 0.7$~yrs after the explosion. Fig.~\ref{fig:SNR_comp} illustrates the density structure in some characteristic snapshots of  such a SNR evolution (hereafter $\rm SNR_{wind}$ model) while Fig. \ref{fig:SNR_dynamics_comp} shows the time evolution of its dynamical properties (shock's radius, velocity and expansion parameter). For comparison we over-plot the SNR model in which the remnant is  evolving in a homogeneous medium of  $n_{ism} = 0.13 ~\rm{cm^{-3}}$ from the start  (hereafter $\rm SNR_{ISM}$ model).

At the beginning of the SNR evolution,  the forward shock of the $\rm SNR_{wind}$ model encounters the dense freely expanding wind region resulting in a denser structure of both the shocked ejecta and ambient medium shells while the radius of the remnant is smaller as compared to the $\rm SNR_{ISM}$ model (see Fig. \ref{fig:SNR_comp} at t= 51 yr). This is also reflected at the FS velocity where for the $\rm SNR_{wind}$ model it remains, during this phase, lower than that of the $\rm SNR_{ISM}$ model. However, as the upstream density of the $\rm SNR_{wind}$ model decreases with the radius, the deceleration of the remnant is less than this of the $\rm SNR_{ISM}$ model and as a result the expansion parameter of the first has higher values than these of the latter (Fig. \ref{fig:SNR_dynamics_comp}). 

About 80 years after the explosion, the forward shock of the $\rm SNR_{wind}$ collides with the shocked wind shell and both its velocity and expansion parameter drop sharply and remain low during the whole phase in which the forward shock propagates in the dense wind shell (Fig. \ref{fig:SNR_dynamics_comp}). This collision  forms a pair of transmitted - reflected  shocks  that move outwards and inwards,  respectively, in a Lagrangian sense (see Fig. \ref{fig:SNR_comp} at t= 101 yr; see also \citet{dwarkadas05} for similar simulations and \citet{Sgro1975} for an analytical approach of this physical process). While the transmitted shock starts to penetrate the shocked wind shell, the reflected shock is moving inwards in a Lagrangian sense  and encounters  the contact discontinuity of the SNR. The collision of the reflected shock with the high density region of the CD  results  in the formation of a second pair of transmitted - reflected  shocks (Fig. \ref{fig:SNR_comp}, at t= 135 yr). The second reflected shock that has been formed by the latter collision is moving now outwards,  and it will collide with the wind shell which now has been (re)shocked by the forward shock of the SNR, and again a third pair of reflected--transmitted shocks will be formed. This process will be repeated several times during the SNR evolution where every time that a reflected shock will collide with the density wall at the contact discontinuity or the shocked wind shell, an extra pair of transmitted-reflected shocks will be formed. The resulting SNR structure will be substantially affected  by this sequence of transmitted and reflected shocks' formation and evolution (Fig. \ref{fig:SNR_comp}, at $t\ge506$ yr). Also note that the small humps depicted in the velocity and expansion parameter plots (more clearly depicted at the latter) correspond to the moments where a fast moving transmitted shock penetrates the whole shocked ambient medium shell and drives the SNR evolution.

When the $\rm{SNR_{wind}}$'s forward shock  fully penetrates the wind bubble shell and starts to propagate in the unperturbed AM ($t \sim 160 $~yr), it accelerates as it encounters the low upstream density of the `new' medium while the pressure behind the shock remains high due to the high density wind material that has been swept up.  Consequently, both the velocity and expansion parameter of the forward shock are higher than these of the  $\rm{SNR_{ISM}}$ model (Fig. \ref{fig:SNR_dynamics_comp}).

Finally, around 600~yr after the explosion the two SNR models seem to reach similar dynamical properties of the forward shock, having almost identical velocity while the expansion parameter of the $\rm{SNR_{wind}}$ model  is slightly higher as the radius of this remnant remains marginally smaller. That means that when the mass of the constant-density medium that is swept up becomes larger than the mass contained in the wind bubble,  the dynamical properties of the forward shock  carry little information about the collision history of the supernova ejecta with the circumstellar bubble.  In our simulations this happens at a radius of 2--3 times the size of the wind bubble.

This is also depicted in the density structure of the two SNRs (Fig. \ref{fig:SNR_comp}). The shocked ambient medium shell of the  $\rm{SNR_{wind}}$ 506~years after the explosion is slightly denser than this of the  $\rm{SNR_{ISM}}$ but these shells get almost  identical structures and densities
as the SNRs evolves further (Fig. \ref{fig:SNR_comp} at $t= 1350$ and 2025 yr). The only difference between the shocked ambient medium shells for the two models  is the density enhancement close to the  contact discontinuity of the  $\rm{SNR_{wind}}$ model.  This is the shell of the wind bubble that has been swept up  by the SNR (see e.g.  Fig.  \ref{fig:SNR_comp} at 2025 yr for $R \sim 6.5 -7.0$ pc). 

 On the other hand, even during the last phase of the simulations ($t= 2025$~yr) the shocked-ejecta shell of the  $\rm{SNR_{wind}}$  model
reveals a completely different structure than the shocked-ejecta shell of the  $\rm{SNR_{ISM}}$ model: the shocked-ejecta shell of the $\rm{SNR_{wind}}$ has a much higher density in the region close to the contact discontinuity, but the density drops faster as we are moving towards the reverse shock, leaving the shocked-ejecta shell thinner as compared to its  $\rm{SNR_{ISM}}$  counterpart.   It is expected that the two SNRs will reveal substantially different emission properties, given that the shocked ejecta are dominating the thermal X-ray emission, and that the emissivity and ionization timescales are strong functions of the density.

%Figures array of the dynamics of a  SNR evolution in a wind bubble as compared to an ISM case

\subsection{Application to Tycho's SNR}\label{sec: appl_Tycho}

 As discussed in the introduction, the ambient medium density estimates based on either dynamical or X-ray emission
 properties of Tycho show discrepancies that may be reconciled if a more complex CSM structure is assumed for Tycho.

Before setting up hydrodynamical models for SNRs evolving in a CSM modified by an outflow from the
 progenitor system, we need to constrain the properties of the CSM as it may have been prior to the explosion of SN1572. Specifically, we have to consider a) what could be the size and the density of the wind bubble and b)  what is the current position of the SNR's forward shock with respect to the pre-existing wind bubble?

The observed value of the expansion parameter  \citep[$m \sim 0.52$ on average and $\sim 0.59$ for the SWW region,][]{katsuda10} cannot be reproduced if currently the SNR's forward shock is in the free-expansion region of the wind, or if it is interacting with the dense wind shell. In the first case,  $m$ would be larger than $0.67$, the transition value to the Sedov-Taylor stage for a SNR in a wind medium \citep{chevalier82}, while if the forward shock would be in the shocked wind/ISM shell a lower expansion parameter would be expected, with a  value that depends on the wind and ISM properties \citep[Fig.  \ref{fig:SNR_dynamics_comp}, c.f.][]{dwarkadas05}.

On the other hand, as shown in Sect.~ \ref{sec: SNR_wind}, when the radius of the SNR is at least  2-3 times larger than the size of the wind bubble the forward shock's expansion parameter approaches  the value that it would have had if the remnant had evolved without a stellar wind bubble. Since the best fit for the expansion parameter in a pure ISM required low density, the same applies for the ISM beyond the wind bubble. 
Combining these two requirements, we find that a wind bubble of a size $< 1.8~d_3$ pc, where $d_3$ is the distance of Tycho in units
of 3 kpc, embedded in an ambient medium with density similar to $0.1~\rm{cm^{-3}}$ does reproduce the dynamics.

The high ionization timescales in Tycho's X-ray spectra requires high shocked-ejecta densities. This can be achieved when the ejecta interacts with high density material early on in the evolution of the SNR. As shown in Fig. \ref{fig:SNR_comp}, an interaction history of the SNR with a dense bubble forms a denser shocked-ejecta shell as compared to that of a SNR in a flat ambient medium. A low density wind cavity works in the opposite direction, as shown by \citet{badenes07}. In order to create a denser bubble in comparison to an ambient medium of density  $\sim 0.1~\rm{cm^{-3}}$,  we 
require that  the density at the outer edge of the free expanding wind bubble is larger than the density of the ambient medium outside the wind region: $\dot{M}/(4 \pi u_{\rm w} r_{\rm w}^2) \gtrsim \rho_{AM}$  where $u_{\rm w}$ the wind terminal velocity, $r_{\rm w}$ the radius of the free expanding wind and $\rho_{\rm AM}$ the density of the unperturbed ambient medium. This requirement constrains the mass loss rate to  $\dot{M} \gtrsim 3\times 10^{-8} \times (u_{\rm w}/10\ {\rm km/s}) \times (r_{\rm w}/{\rm pc})^2~ \rm{M_{\odot}~yr^{-1}}$.

 In conclusion, a small, dense, wind bubble in a low density ISM allows for both the required high ionization age, as well as the observed expansion parameter. 

\section{Numerical models of dynamics and emission properties of Tycho}

In this section we present the result of our simulations that are targeted to model Tycho at its current age ($\sim 440$~yrs). We aim at getting the best parameters for the wind model that is able to explain both dynamics and emission properties, and compare it to the best fit models for SNR evolution in a homogeneous medium.

\subsection{Method}

We use the {\sc AMRVAC}  code (described in Sect. \ref{sec:study}) to simulate the dynamical properties of various models, from which we extract the expansion parameter, as well as the relative positions of the forward and reverse shock and the contact discontinuity. 

In order to build the synthetic spectra of our models, we employ the combination of the hydrodynamical package {\sc SUPREMA} \citep{Sorokina2004} and the X-ray spectra fitting software {\sc SPEX} \citep{Kaastra96}. The {\sc SUPREMA} code solves the system of spherically symmetric differential equations in Lagrangian coordinates and uses an implicit finite-difference method developed by \citet{blinnikov98}. It incorporates a number of physical processes relevant for the shocked medium, e.g. electron thermal conduction, ion-electron temperature equilibration and radiative losses. The code takes into account self-consistently the ionization state of the shocked plasma, and,
for each zone and at every time step,
the kinetic calculations are performed for all species of the most abundant elements.

Recently the package was updated with an interface to the SPEX X-ray spectral fitting code in order to calculate a detailed X-ray emission spectrum for each zone of the SUPREMNA hydrodynamical grid, taking into account the local non-equillibrium ionization structure. The total X-ray spectrum is calculated by summing all the elements over the entire hydrodynamical mesh. Note that there is no self-consistent theory to predict the X-ray synchrotron radiation
contribution to the spectrum \citep{warren05,cassam07}, so a power law component was separately fitted to the spectrum to fit the X-ray continuum.

 For our modeling we use two SN Ia explosion models: 
a deflagration model, which results
from a thermonuclear, subsonic flame propagation through the white dwarf, and
a delayed detonation (DDT) model , in which the initial, subsonic flame changes into
a shock behind which nucleosynthesis occurs.
For the deflagration model we used the canonical deflagration model
W7 \citep{nomoto84}.
For the delayed detonation model we used the model DDTc \citep{bravo2005,badenes03,badenes05}, 
which up to now seems to be the model that best reproduces the X-ray spectra of Tycho \citep{badenes06}. 
For the AMRVAC code we follow an exponential interpolation of the data points for the ejecta density profile and a linear interpolation for the velocity one. The grid size, of the AMRVAC simulations,  is $2.2 \times 10^{19}$~cm $\approx 7.1$~pc. On the base level, we use 540 cells and allow for 6 refinement levels for the wind bubble formation and 7 levels for the SNR evolution. Hence, the maximum effective resolution is  $1.3 \times 10^{15}$~cm and  $6.4 \times 10^{14}$~cm, respectively. At the {\sc SUPREMA}  simulations   we use, for the case of W7 explosion, 300 Lagrangian zones, 99 for the ejecta and the rest for the ISM or CSM. For DDTc explosion the hydrodynamical mesh comprises 200 zones, 81 of which comprise the ejected supernova material.  We expand the original explosion model up to the age of 10 years and then start the numerical modeling. The supernovae are surrounded by a medium  at rest, with a constant temperature of $10^4$~K, and with solar elemental abundances\footnote{We have verified that the modified chemical composition of the progenitor wind only weakly affects the resulting X-ray spectra and can be ignored.}. We set the degree of the temperature equilibration behind the shocks  of $T_e/T_i \simeq 0.1$. This
modest non-equilibration at the shock front is roughly consistent with observations \citep[e.g.][]{ghavamian07,vanadelsberg08,Broersen2013}.

For each explosion model (W7, DDTc) we study three different cases for the medium that surrounds the explosion center.  First we consider that the SNR is embedded  in a homogeneous ambient medium. Treating the ambient medium density as a free parameter we find the two values that provide the best agreement with a) the dynamics (hereafter ${\rm SNR_{ISM,dyn}} $ model), or b) the X-ray spectra (hereafter ${\rm SNR_{ISM,spectra}}$ model) of Tycho SNR. Then for the third case, c)  we introduce the wind bubble and we let the SNR interact with it  (hereafter ${\rm SNR_{wind}}$ model).  The wind properties are chosen in such a way  that, on the one hand, the dynamics of the SNR is in agreement with the current observations of Tycho but,
on the other hand, that the density of the shocked ejecta -at the age of Tycho- is similar to the ${\rm SNR_{ISM,spectra}}$  model. 
Thus, we ensure that the `best-fit' ionization structure of the shocked ejecta  is preserved.  This required running several models in order to find the best solutions. Here we only present the optimal solution.

 We compare the results of our simulations with observations of the southwest-west (SWW) region of Tycho ($\rm 200^o - 300^o$ counterclockwise starting from the north, see Fig. \ref{fig:mos-sky}). 
 The reason for using that region is twofold. First of all the shock front in this region appears smoother than the other 
 regions and the expansion parameter, in the SWW, shows less variation than in the northwestern region \citep{reynoso97,katsuda10}.
Recently,  \citet{williams13} showed that the densities in the north/northeastern region are also higher than this in the SWW.
These observational results support the idea of \citet{reynoso99} and \citet{lee04} that in the north/northeastern the shock wave is interacting with a molecular cloud. However, this idea is not universally accepted \citep[e.g.][]{tian11}.  In the case of interaction with molecular cloud, or
with a local density enhancement,  the deviations from sphericity and variation in expansion rate are likely caused
 by accidental features of the local ISM, whereas in the present paper we are interested in the CSM structure that is related directly to outflows from
 SN Ia progenitor. The second reason to compare our results only to the SWW region is that this allows for a direct comparison
 with the models of  \citet{badenes06}, who also focussed, for very much the same reasons, on this region.  Thus, hereafter all of the observational data (spectra, expansion parameter, relevant positions of the FS, CD, RS) are averaged or spatially integrated over this specific region.   We extract the forward shock expansion parameter and the ratios of the contact discontinuity and reverse shock over the forward shock (CD:FS, RS:FS) from the AMRVAC code, and compare them with the values measured by \citet{reynoso97}, \citet{katsuda10} (for the expansion parameter) and \citet{warren05} (for CD:FS, RS:FS). Finally, we compare the results of our synthetic X-ray spectra with the observed X-ray spectra of Tycho using the XMM Newton observations of the remnant (OBS ID~0096210101) \footnote{The EPIC MOS data were processed with the pipeline routine for analysis of the extended object, presented by \citet{Snowden2011}.}.

\subsection{Results}

%\begin{sideways}
%\begin{figure*}%[htbp]    & trim l b r t
\begin{figure*}%[htbp]    & trim l b r t
%\begin{center}
\begin{tabular}{ccc}

\includegraphics[trim=26 15 16 22,clip=true,width=55mm,angle=0]{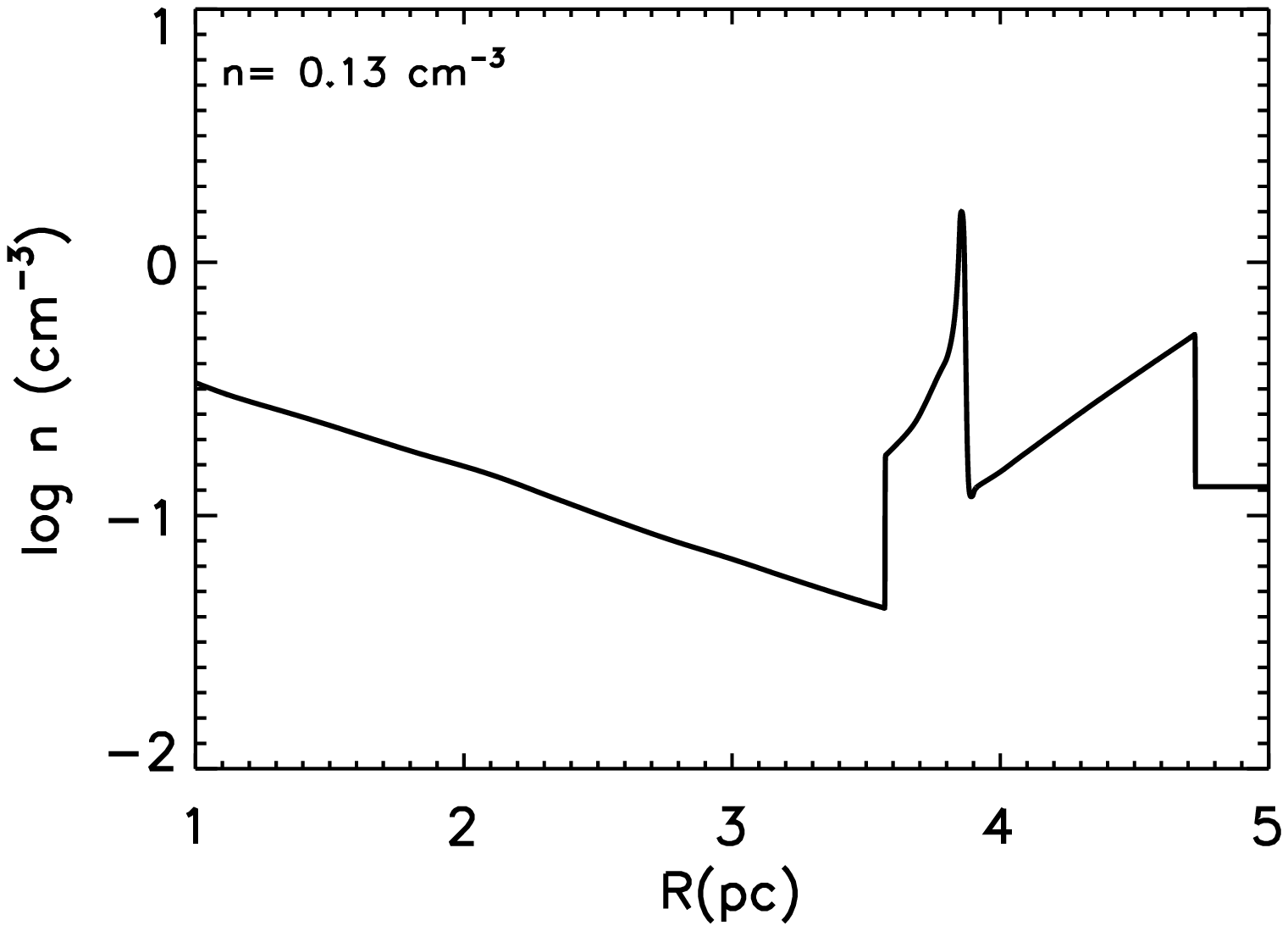} 
\includegraphics[trim=26 15 16 22,clip=true,width=55mm,angle=0]{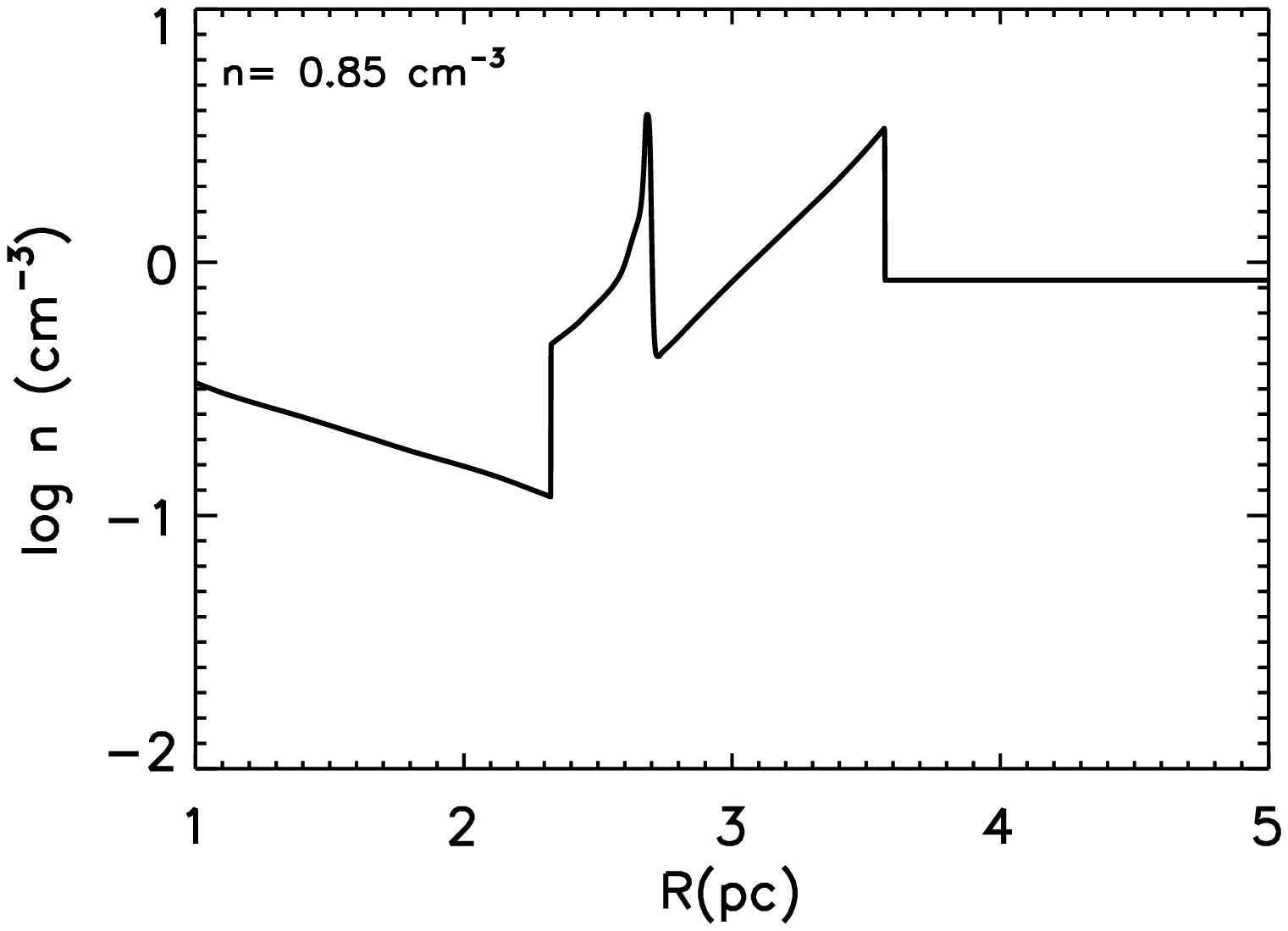} 
\includegraphics[trim=26 15 20 22,clip=true,width=55mm,angle=0]{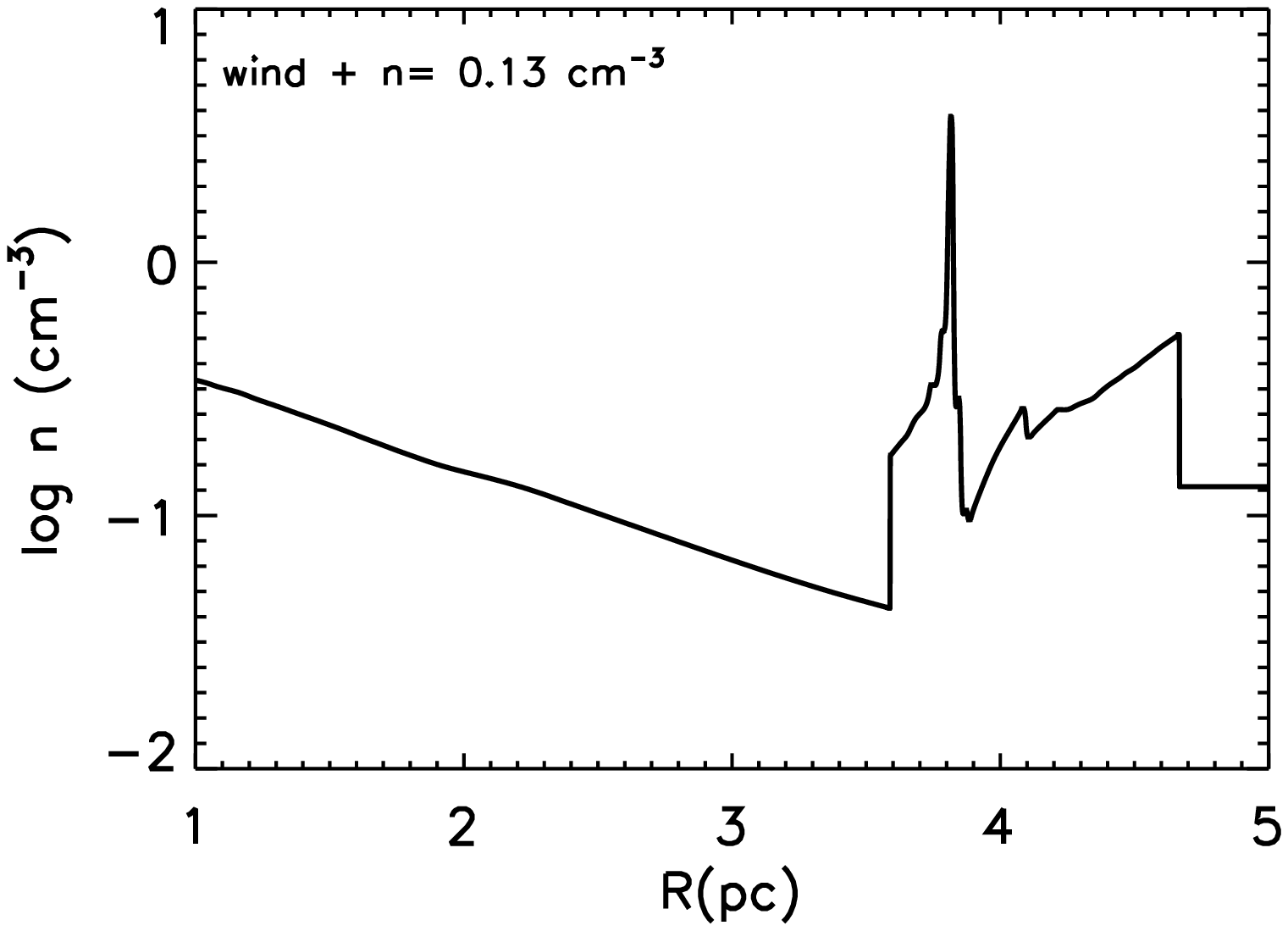} 
\\
\\
\includegraphics[trim=26 15 16 22,clip=true,width=55mm,angle=0]{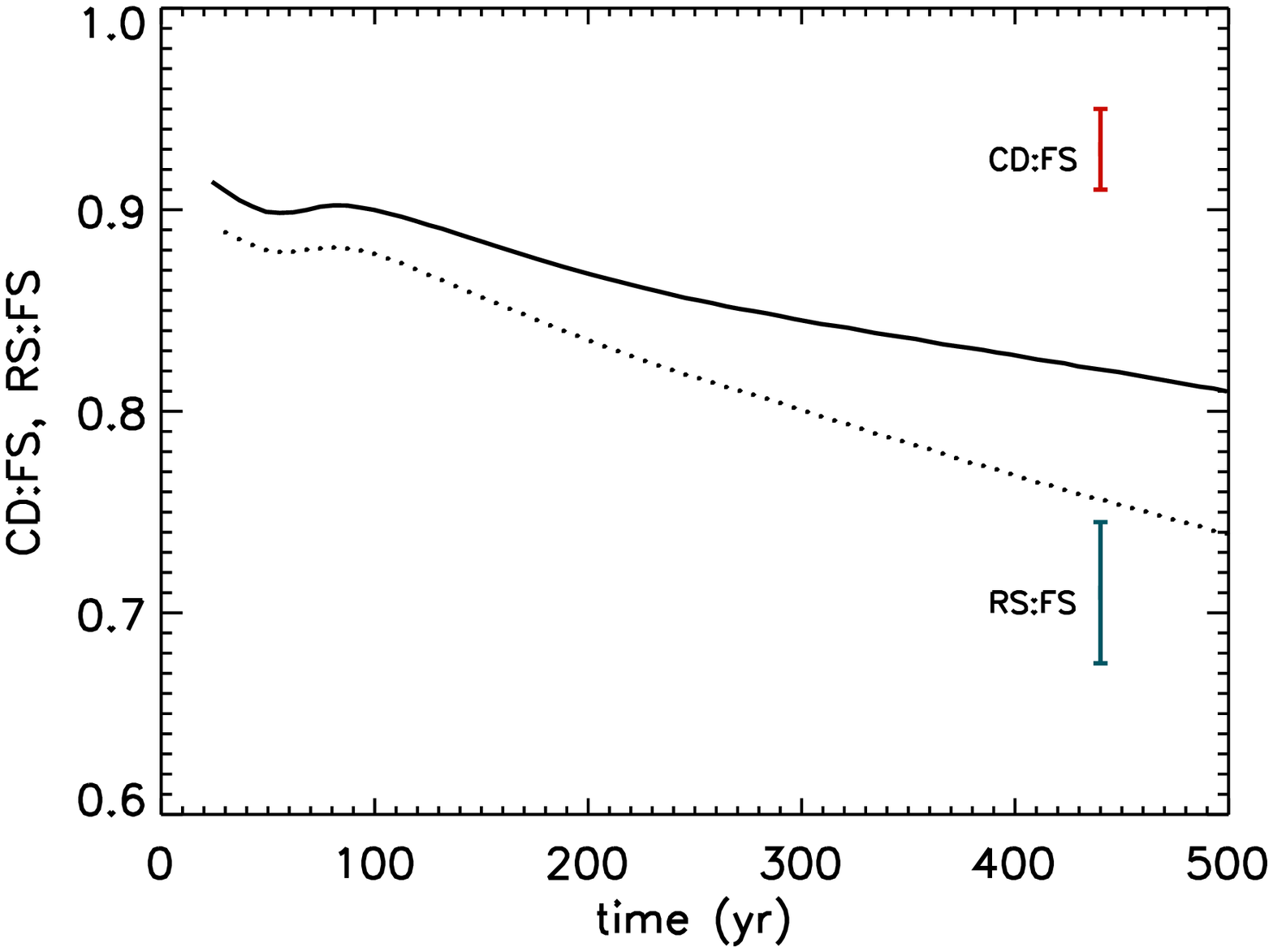} 
\includegraphics[trim=26 15 16 22,clip=true,width=55mm,angle=0]{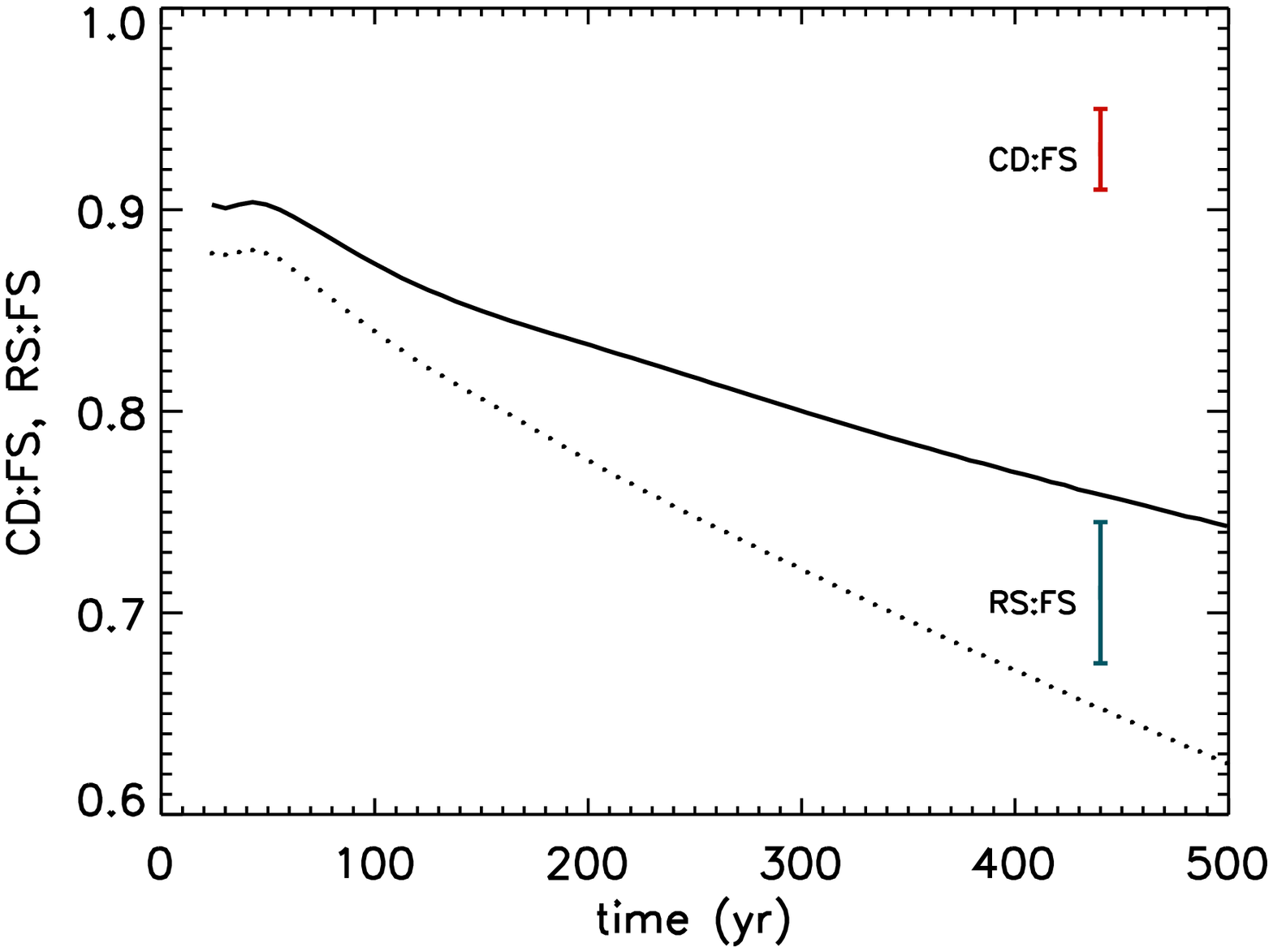} 
\includegraphics[trim=26 15 20 22,clip=true,width=55mm,angle=0]{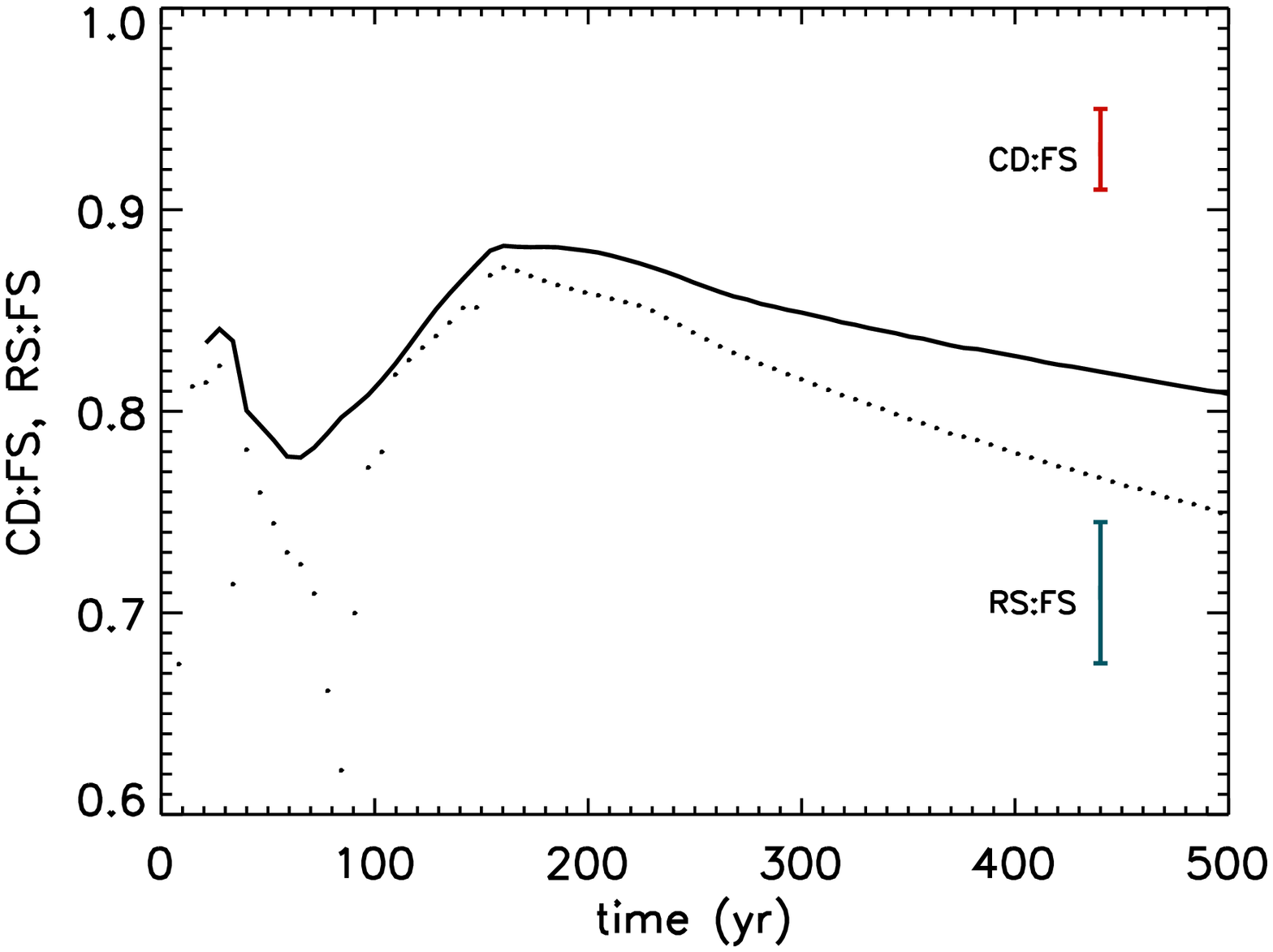} 
\\
\\
\includegraphics[trim=26 15 16 22,clip=true,width=55mm,angle=0]{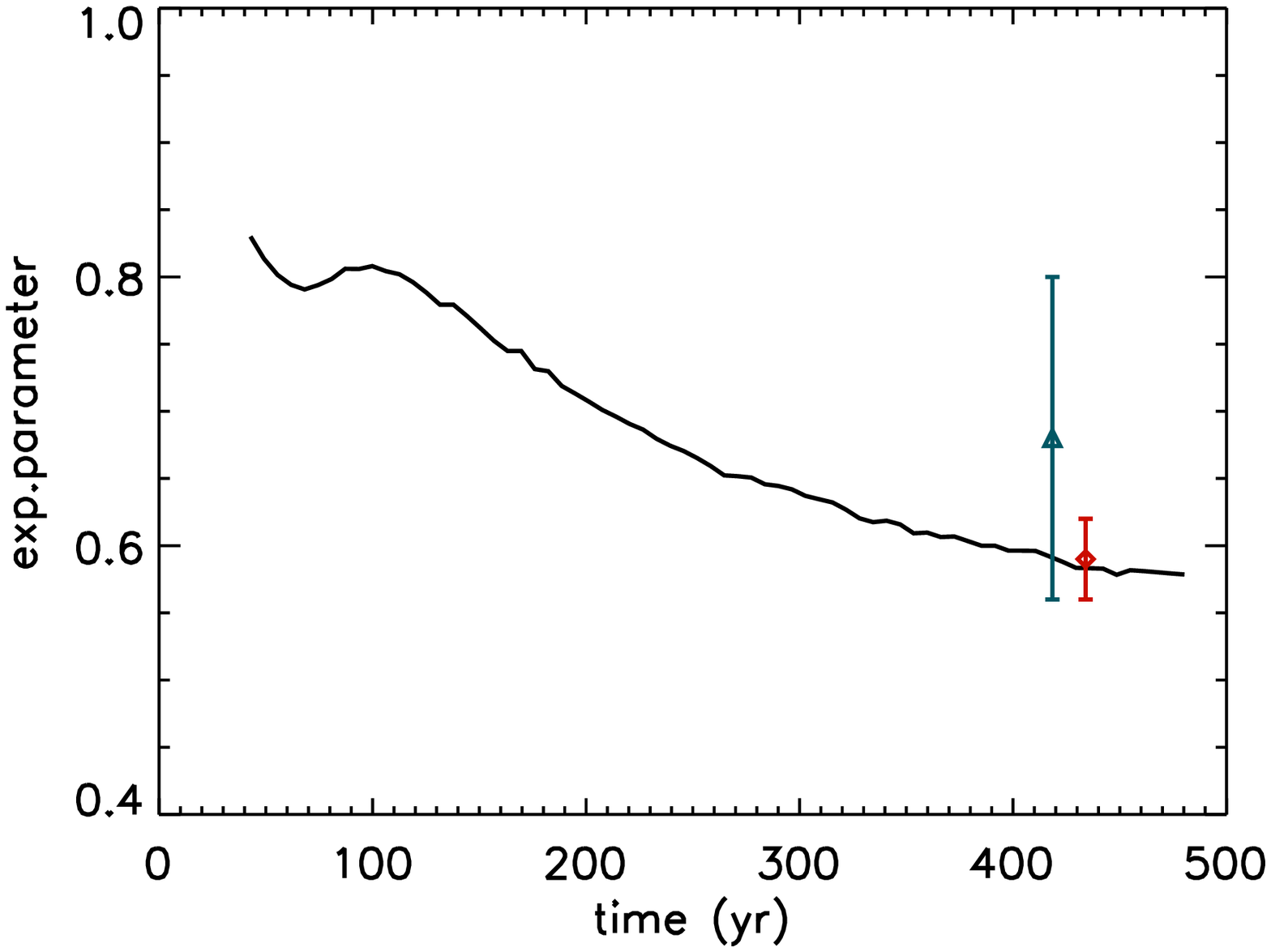} 
\includegraphics[trim=26 15 16 22,clip=true,width=55mm,angle=0]{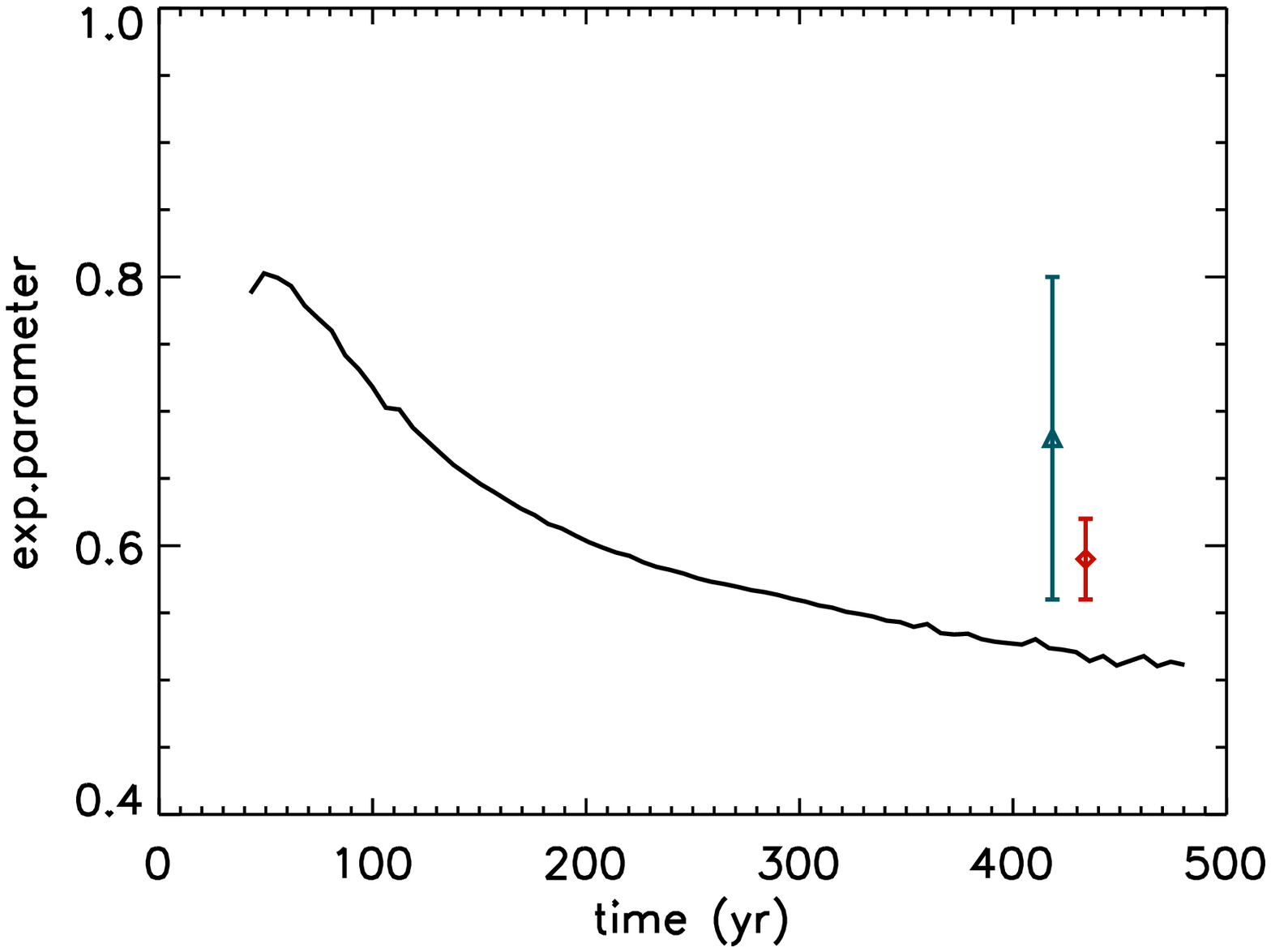} 
\includegraphics[trim=26 15 20 22,clip=true,width=55mm,angle=0]{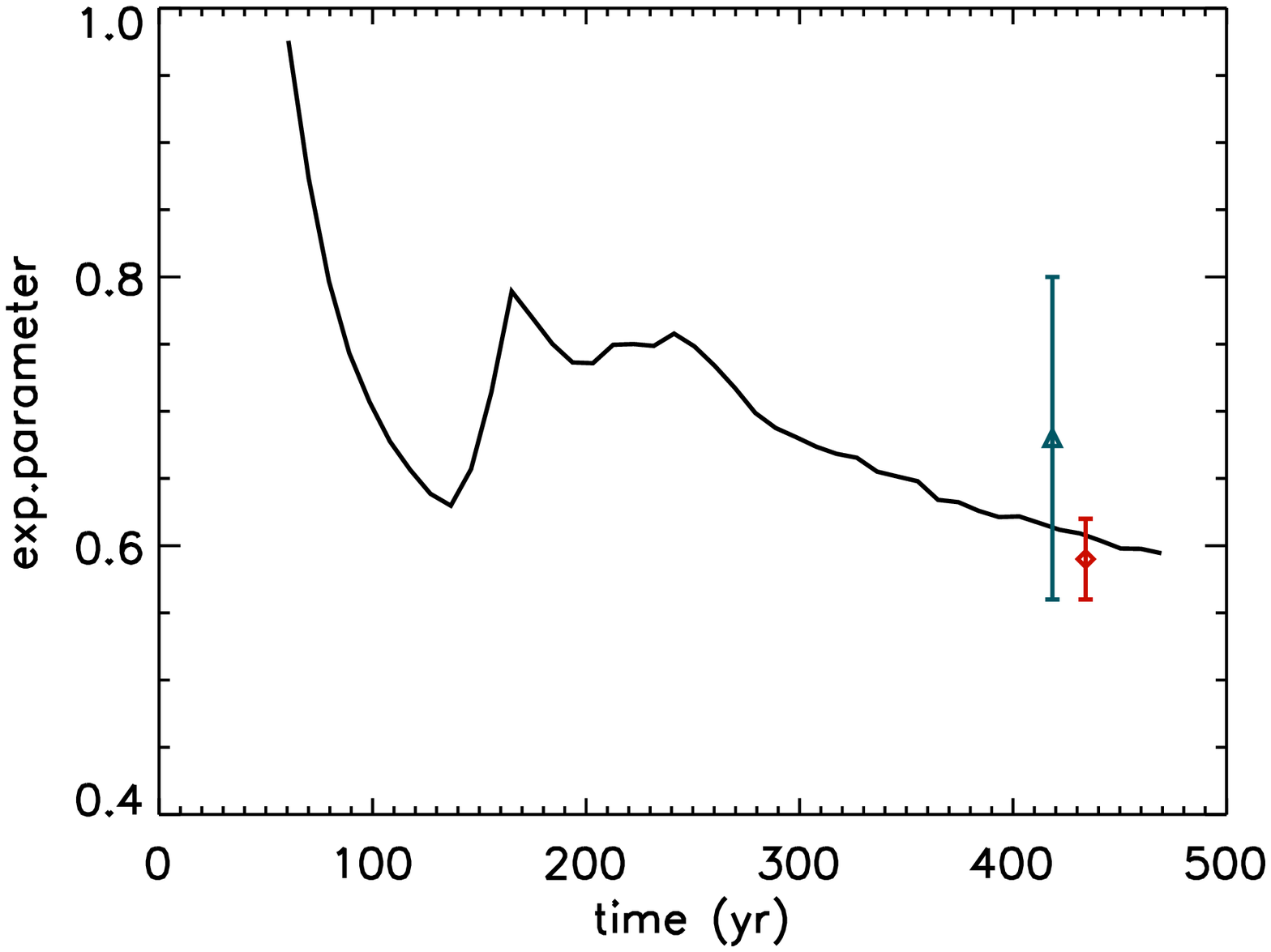}
\\
\\
\includegraphics[trim=36 0 50 22,clip=true,width=0.31\hsize,angle=0]{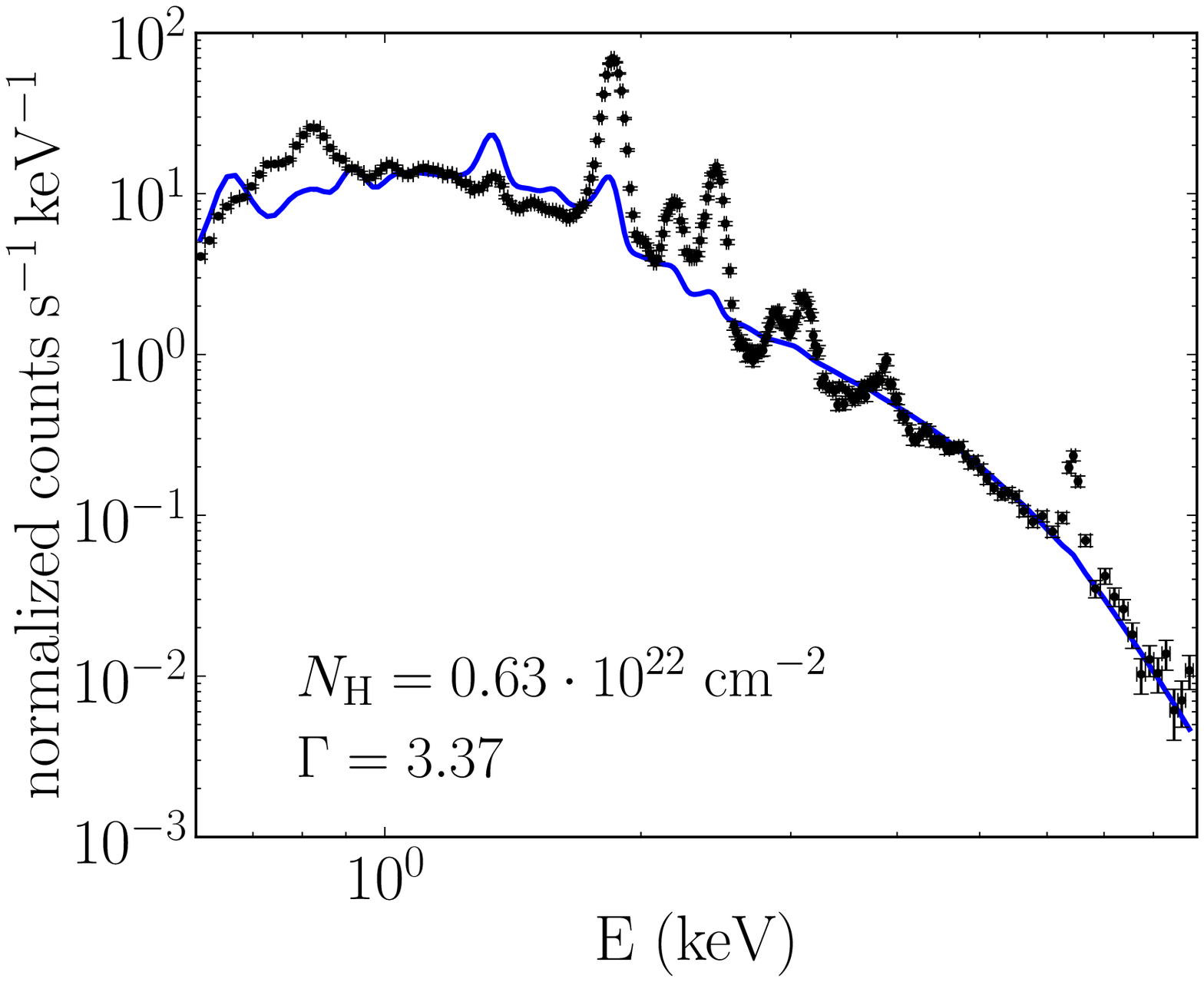}  \hfill
\includegraphics[trim=62 0 50 22,clip=true,width=0.29\hsize,angle=0]{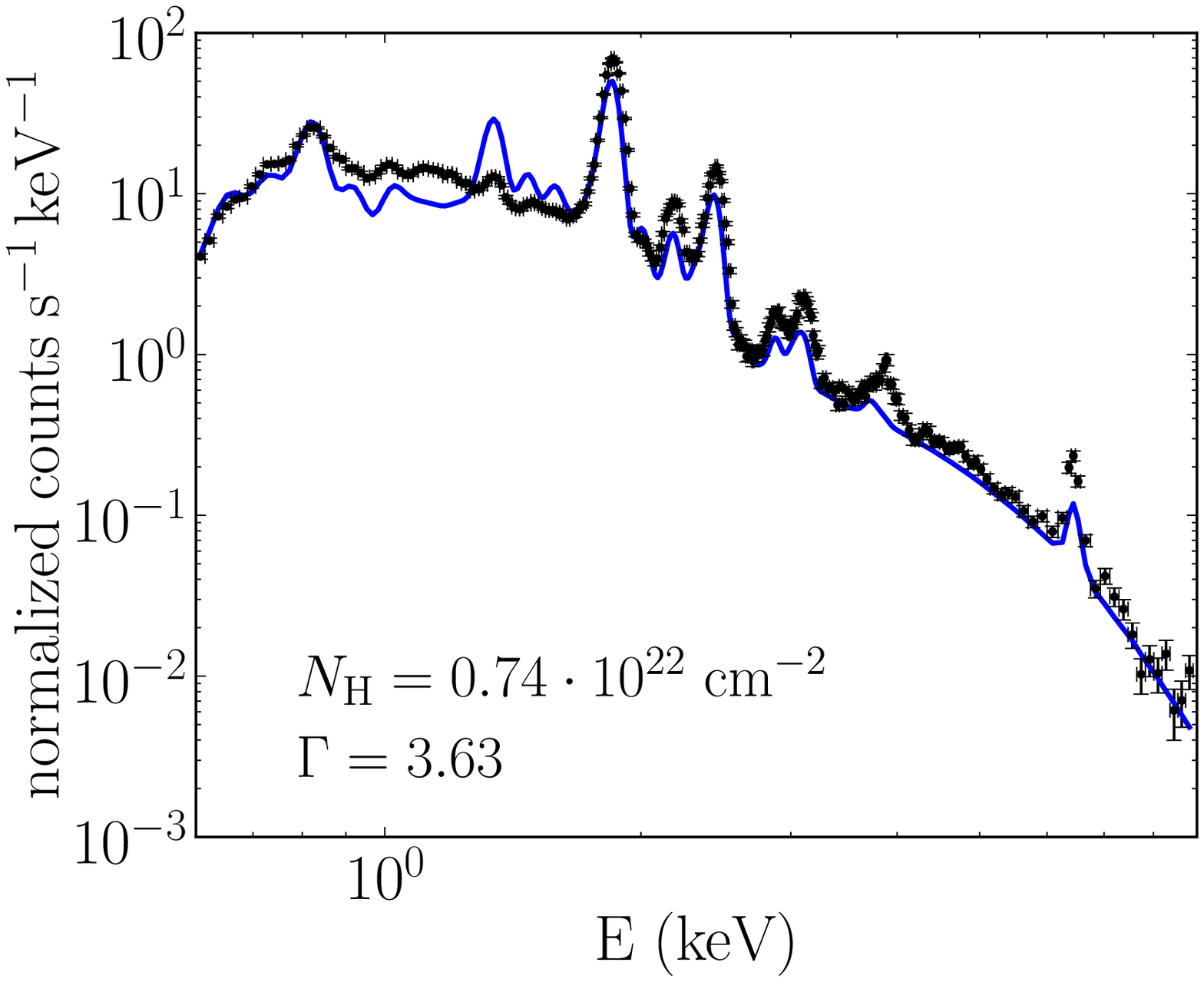} \hfill
\includegraphics[trim=62 0 50 22,clip=true,width=0.29\hsize,angle=0]{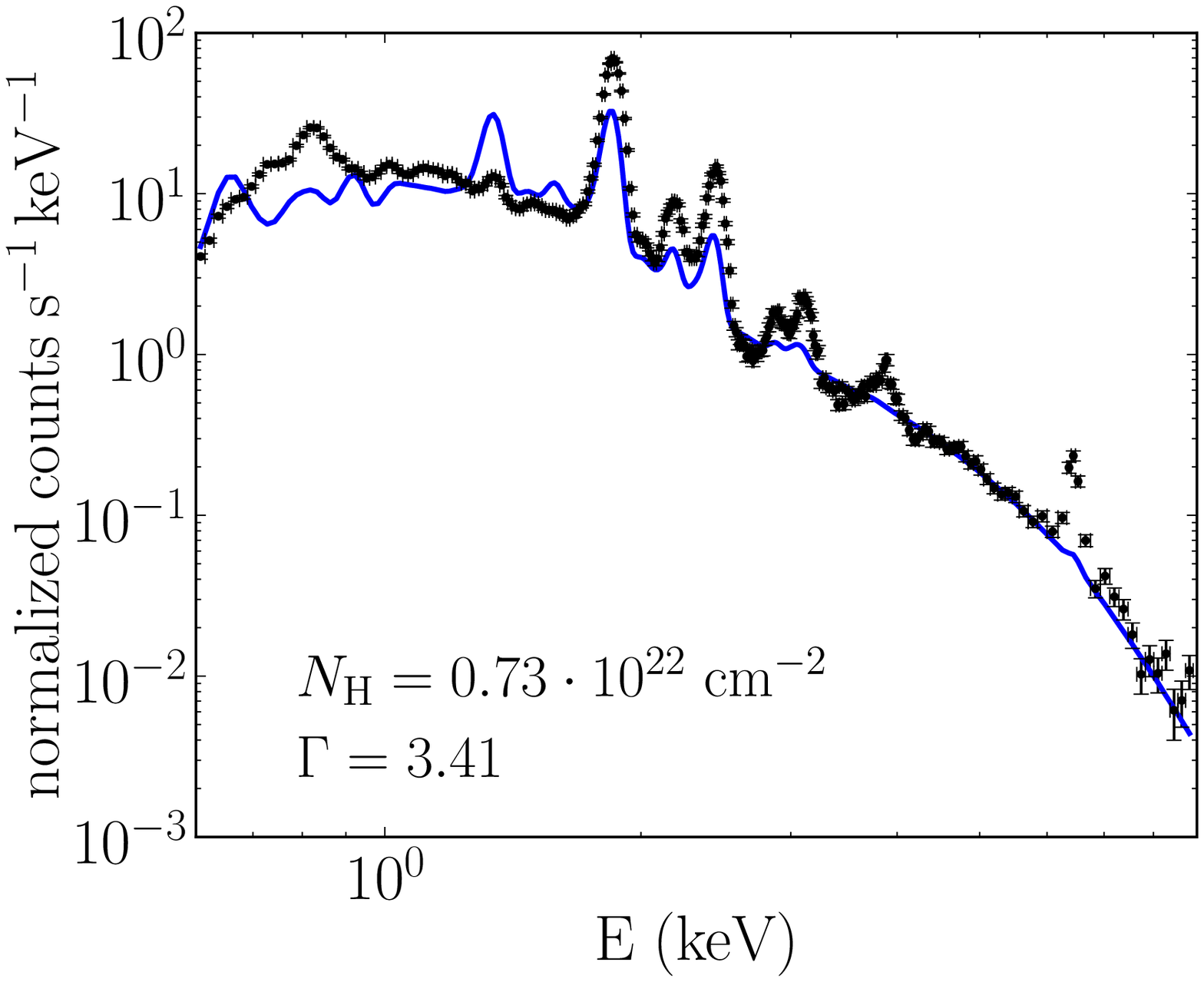} 

\end{tabular}
\caption{Results of the simulations for the W7 explosion model, with different properties of the ambient medium.  
The left column is for a SNR evolving in a flat ISM of density $n=0.13~\rm{cm^{-3}}$ (SNR$_{\rm ISM,dyn}$), the central column is for a SNR evolving in a flat medium of $n=0.85~\rm{cm^{-3}}$ (SNR$_{\rm ISM,spectra}$),
and the right column is a SNR with a wind bubble interaction history (SNR$_{\rm wind}$).
The different rows are, from top to bottom:
 the density structure of the SNR at the current age of Tycho, the ratio of the contact discontinuity (solid line) and of the reverse shock (dotted line) radius over the radius of  forward shock,  the expansion parameter of the forward shock and the resulted spectra (blue solid line) in comparison to the observed data points. The vertical bars at the RS:CD:FS plots refer to the observed values of these ratios \citep{warren05}, while points at the expansion parameter plots refer to the relevant observations performed in the radio \citep[cyan triangle; ][]{reynoso97} and the X-ray \citep[red diamond;][]{katsuda10} band.  The bottom row indicates for each model the best-fit values for the power law slope, $\Gamma$, and hydrogen absorption column density, $N_{\rm H}$. All of the observational data refer to the SWW region of the remnant (see text for details).}
 \label{SN_comp}
\end{figure*}

\begin{figure*}%[htbp]    & trim l b r t
%\begin{center}
\begin{tabular}{ccc}

%\begin{array}{ccc}
\includegraphics[trim=26 15 16 22,clip=true,width=55mm,angle=0]{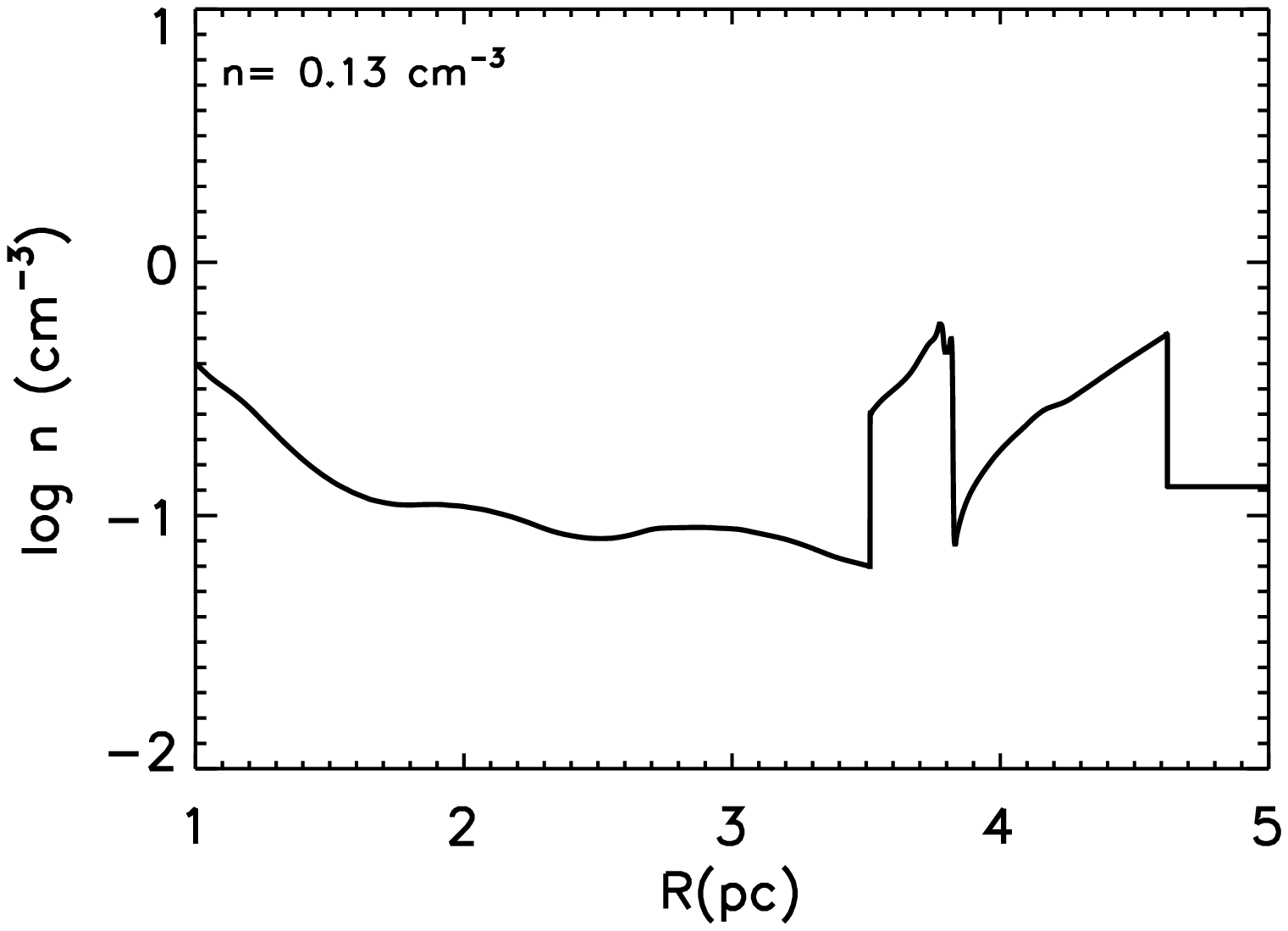} 
\includegraphics[trim=26 15 16 22,clip=true,width=55mm,angle=0]{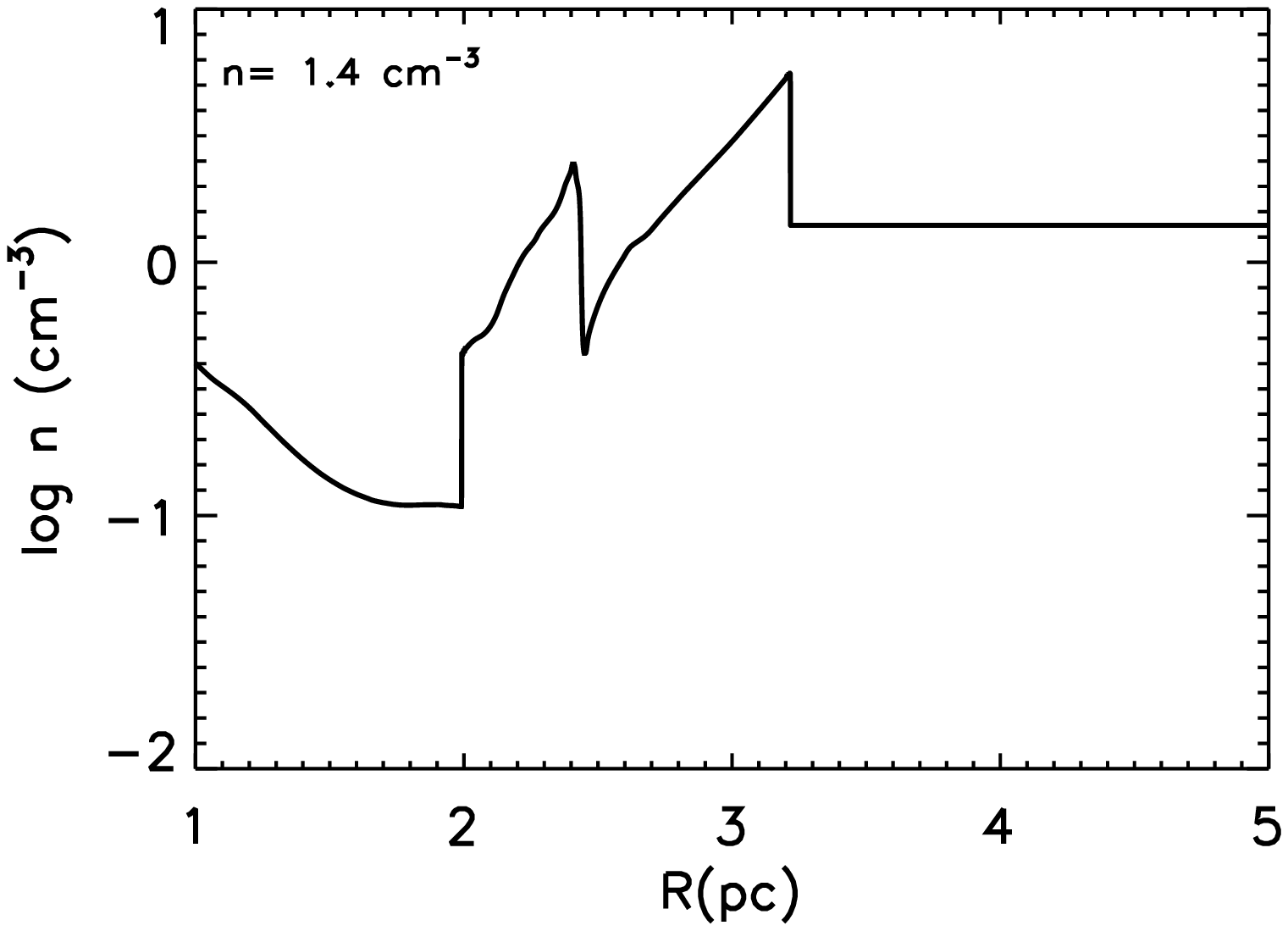} 
\includegraphics[trim=26 15 20 22,clip=true,width=55mm,angle=0]{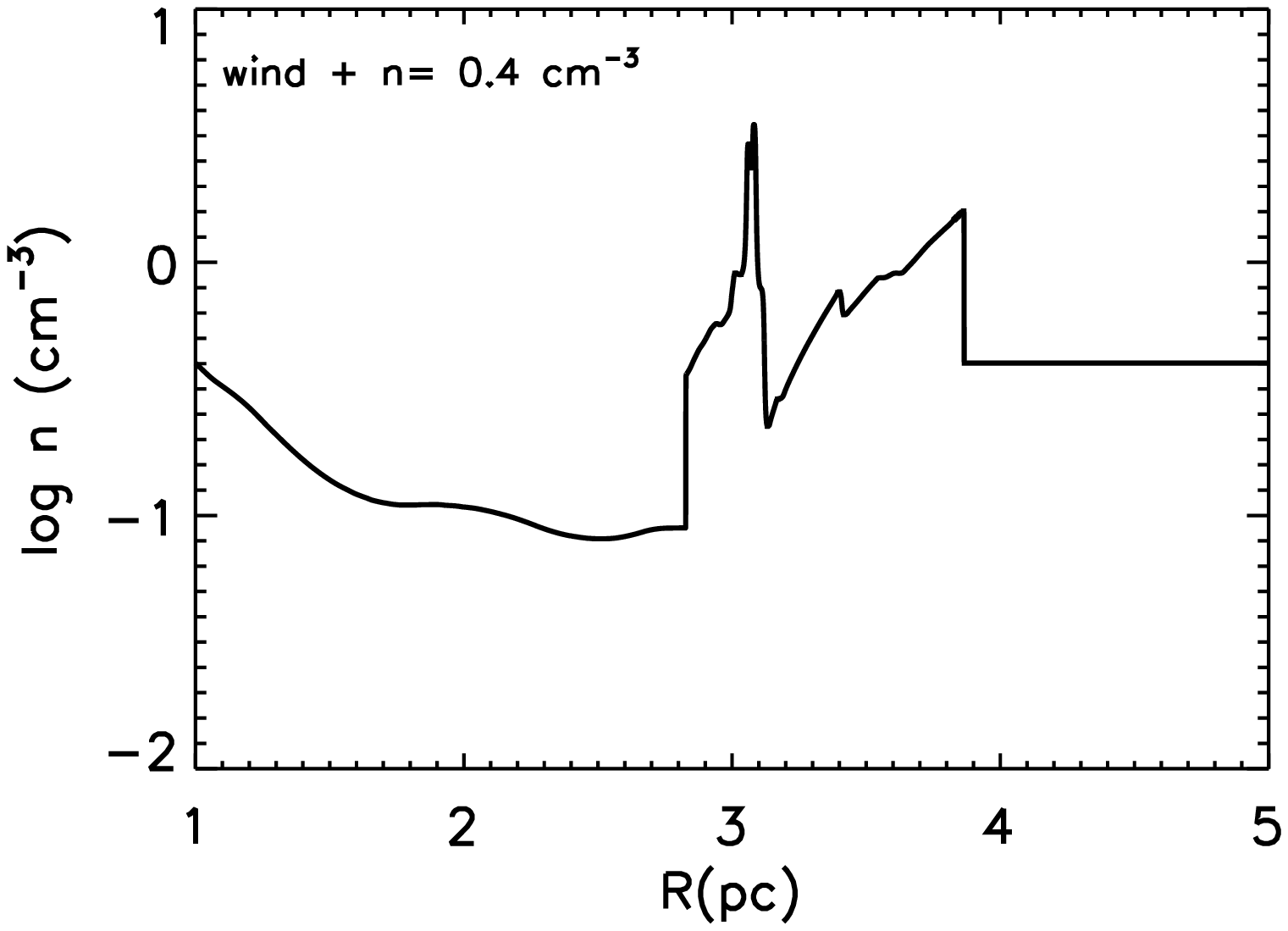} 
\\
\\
\includegraphics[trim=26 15 16 22,clip=true,width=55mm,angle=0]{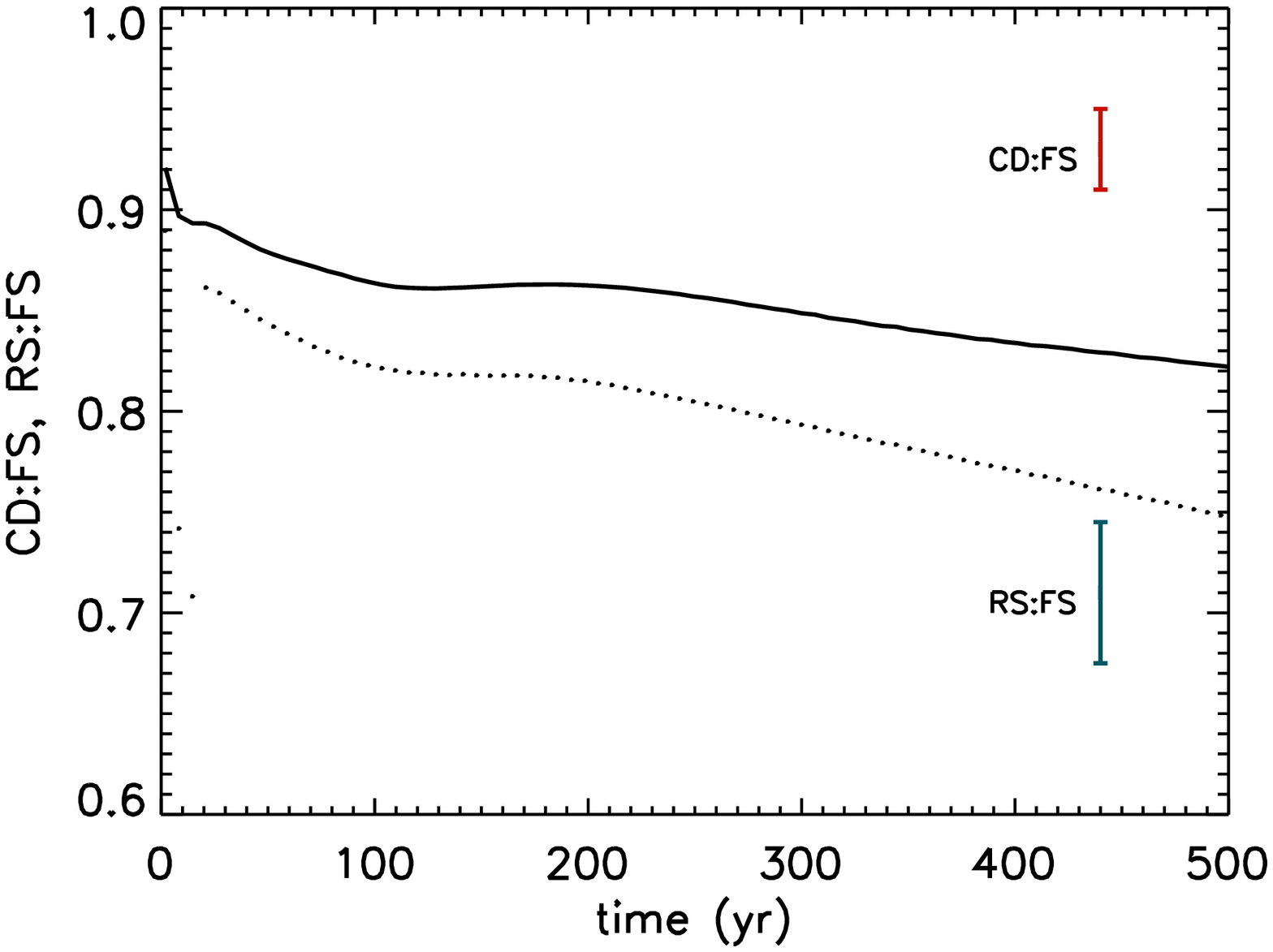} 
\includegraphics[trim=26 15 16 22,clip=true,width=55mm,angle=0]{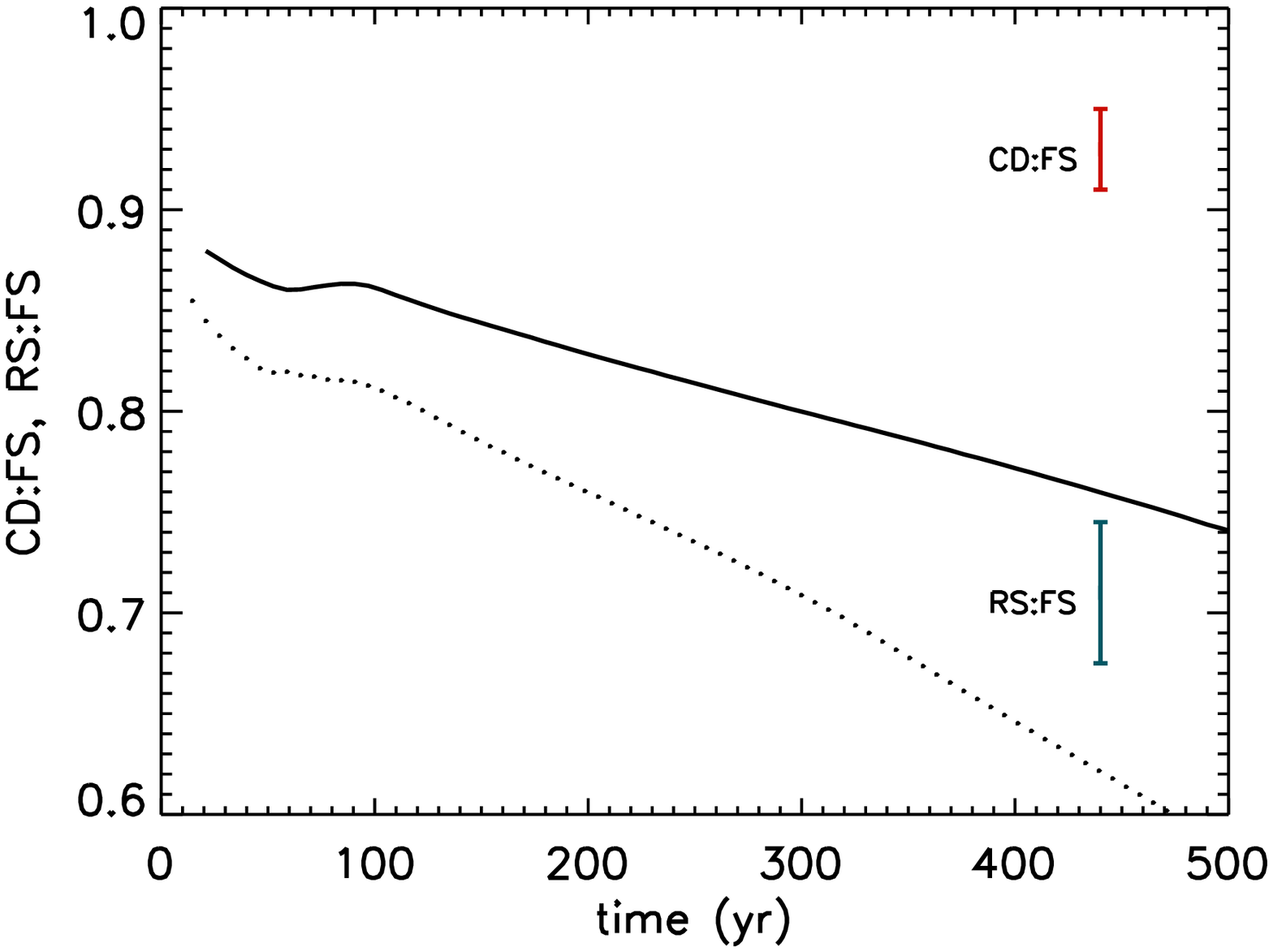} 
\includegraphics[trim=26 15 20 22,clip=true,width=55mm,angle=0]{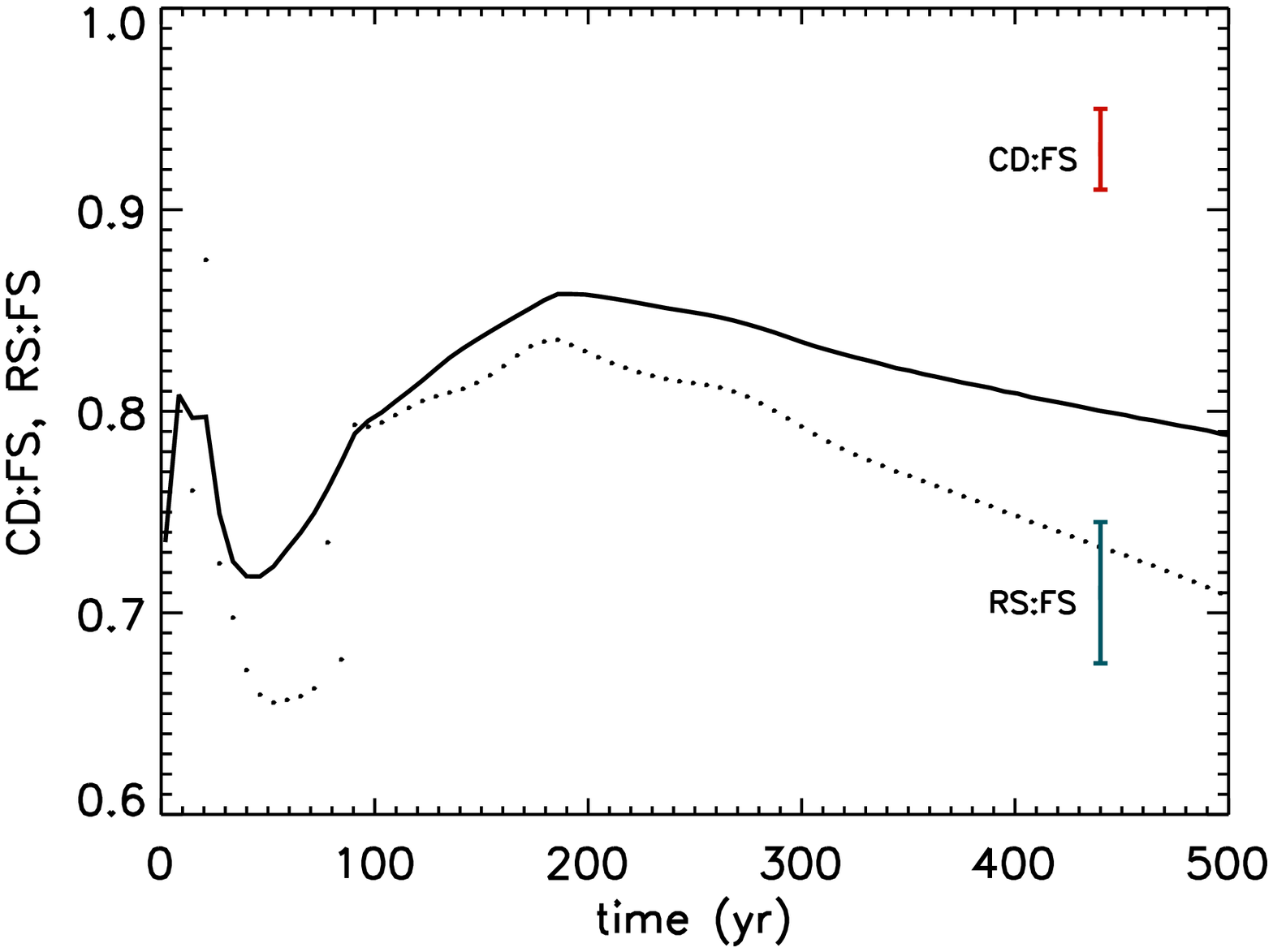} 
\\
\\
\includegraphics[trim=26 15 16 22,clip=true,width=55mm,angle=0]{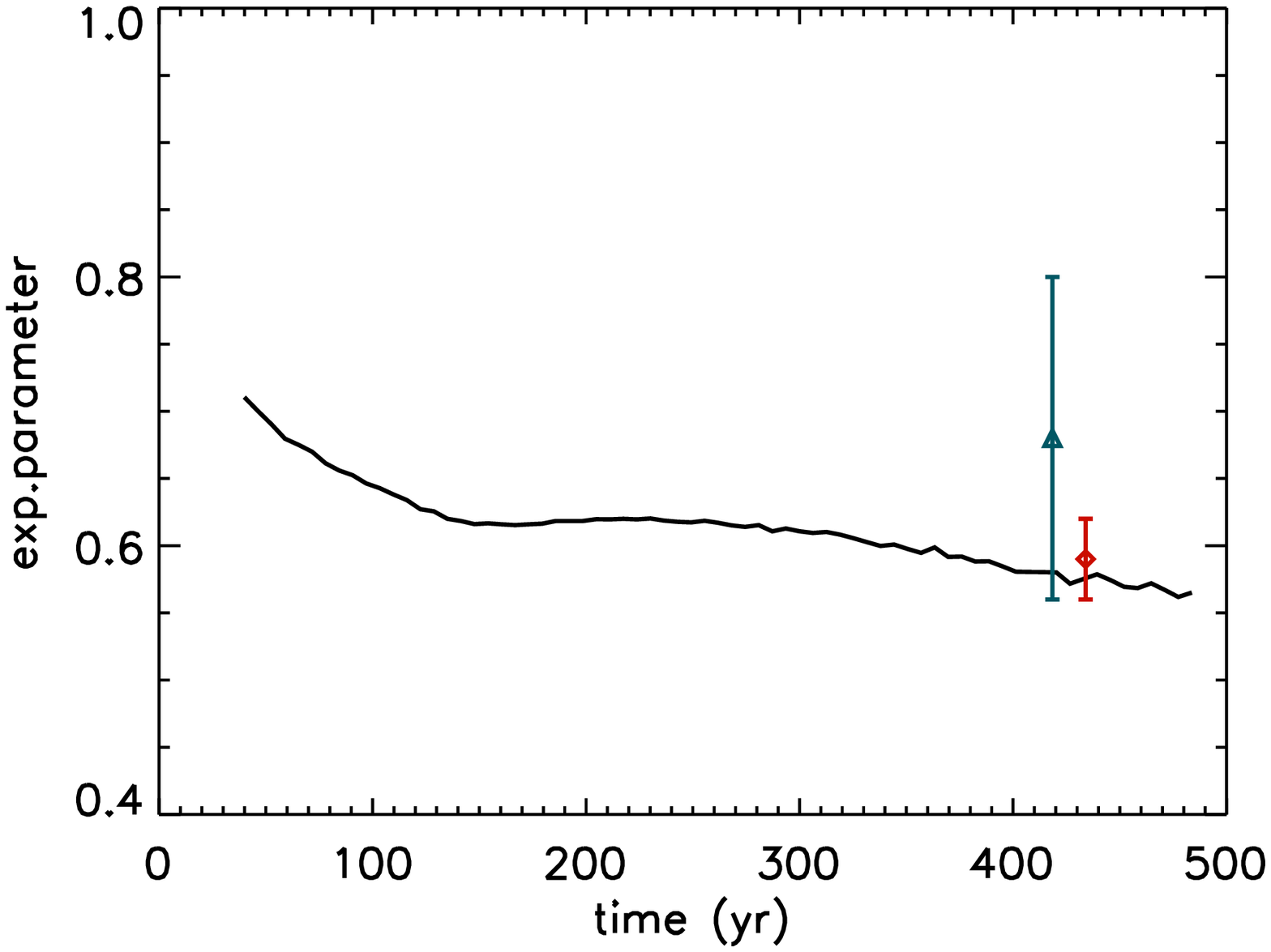} 
\includegraphics[trim=26 15 16 22,clip=true,width=55mm,angle=0]{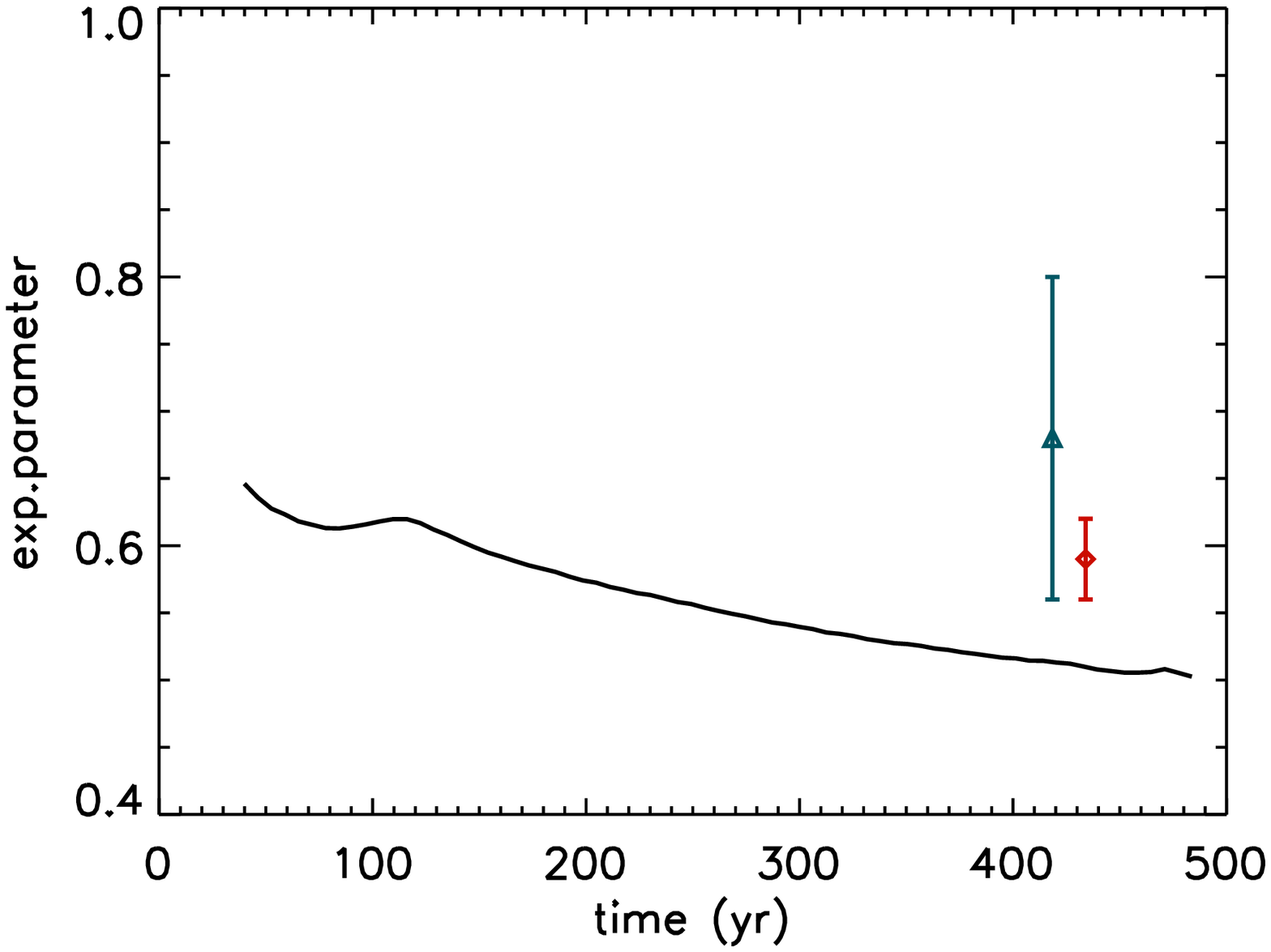} 
\includegraphics[trim=26 15 20 22,clip=true,width=55mm,angle=0]{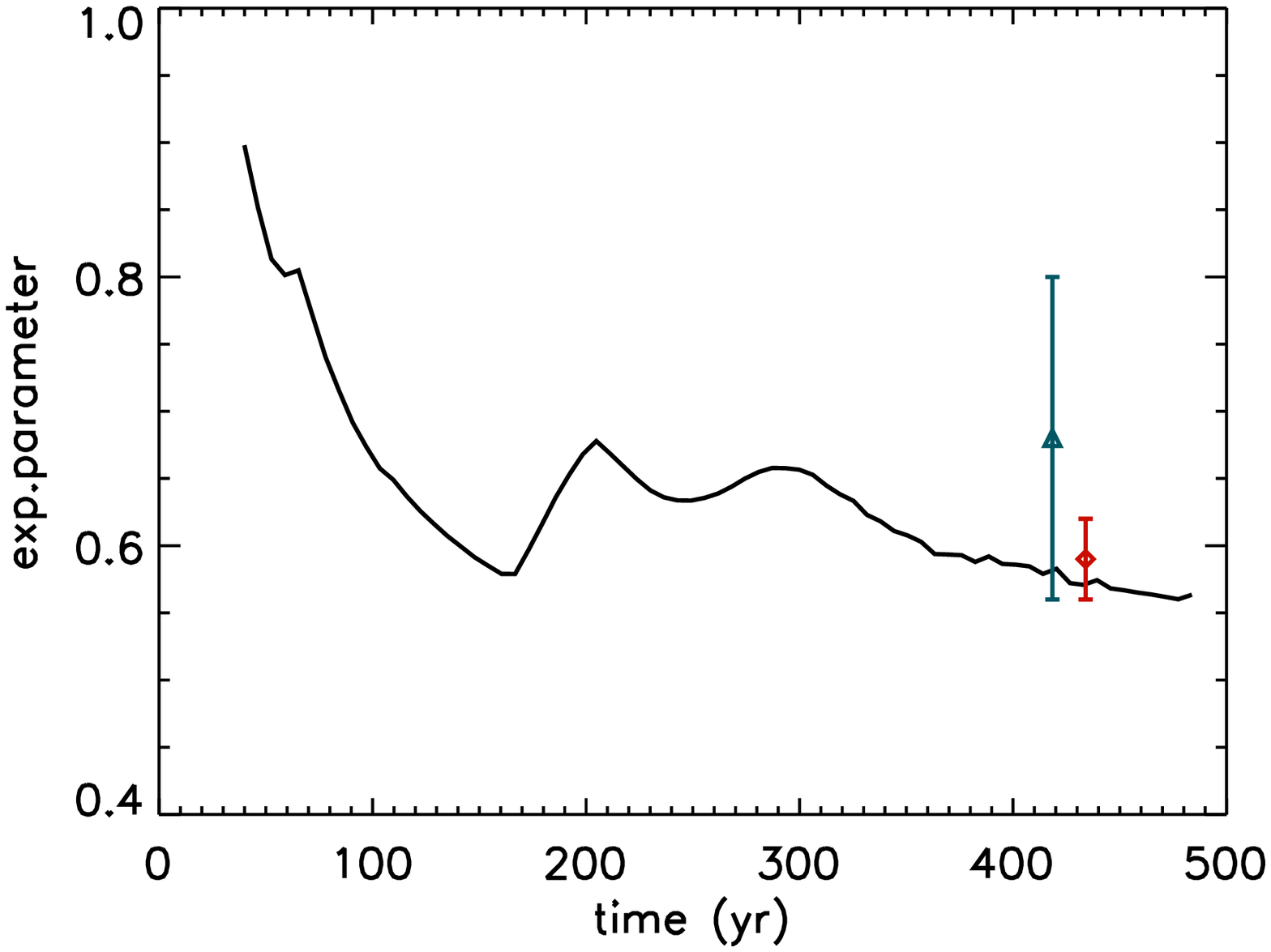}
\\
\\
\includegraphics[trim=36 0 50 22,clip=true,width=0.31\hsize,angle=0]{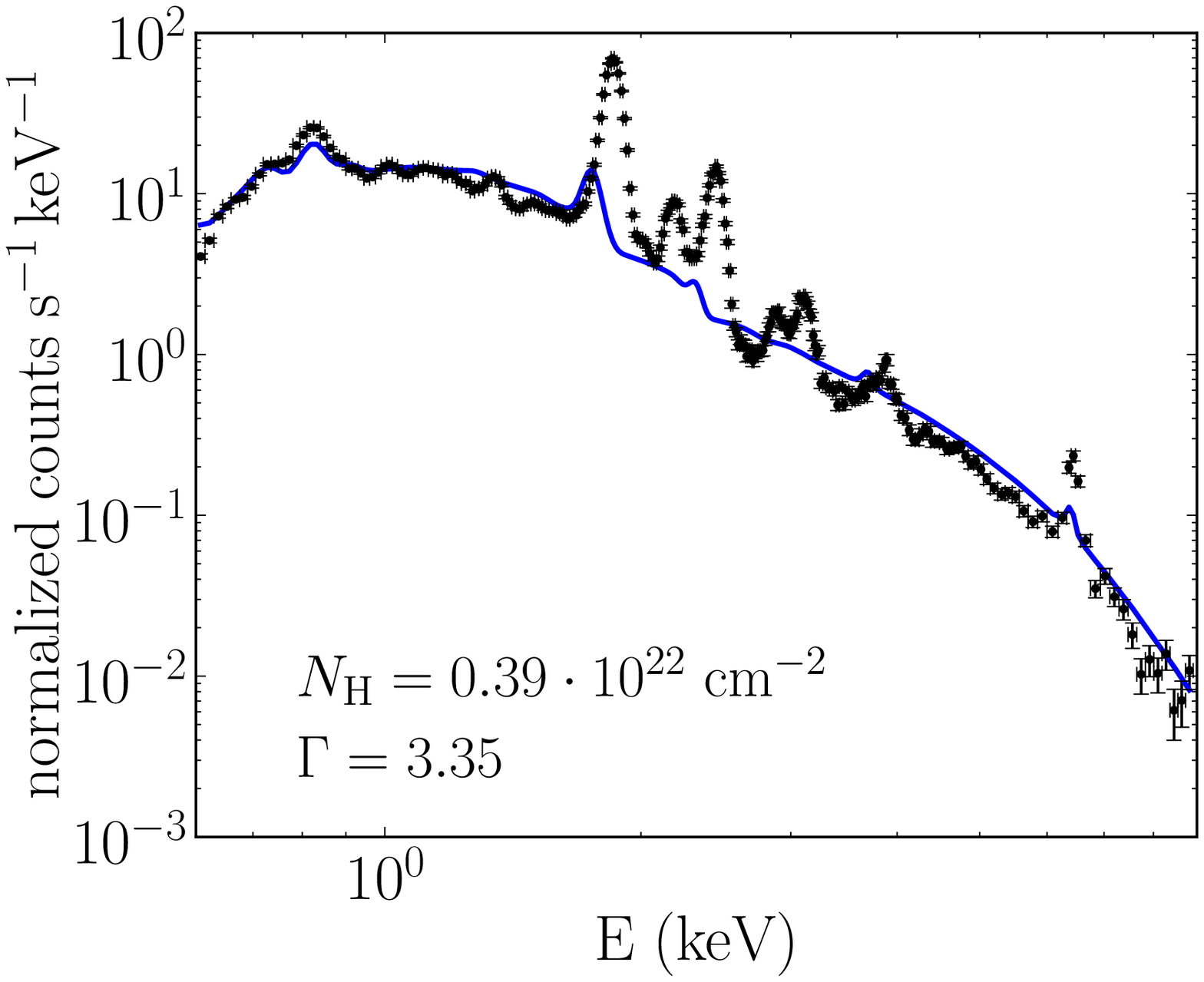} \hfill
\includegraphics[trim=62 0 50 22,clip=true,width=0.29\hsize,angle=0]{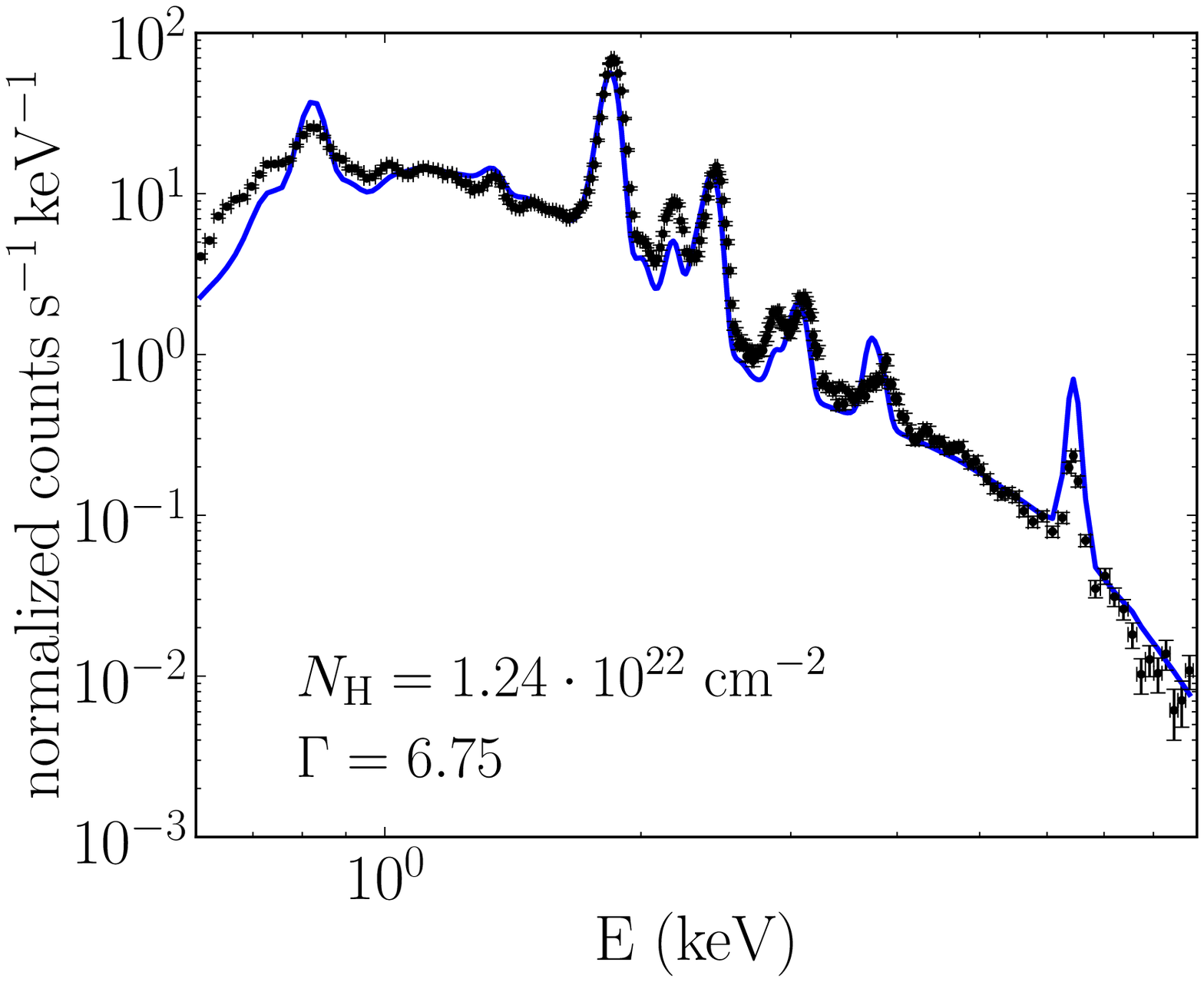} \hfill
\includegraphics[trim=62 0 50 22,clip=true,width=0.29\hsize,angle=0]{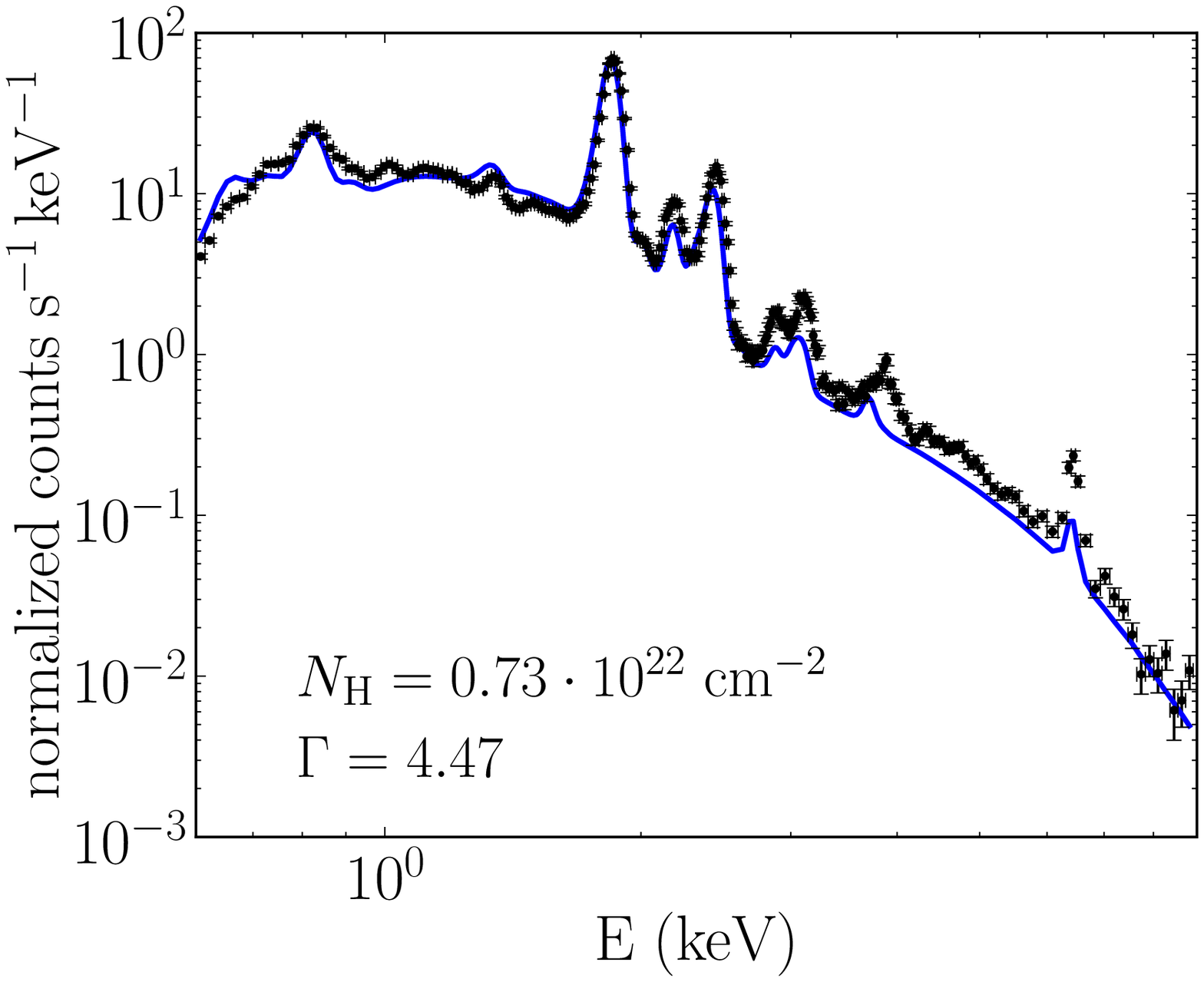}

\end{tabular}
%\end{center}
\caption{.Same as Fig. \ref{SN_comp} but for the DDTc explosion model. }  \label{SN_compDDTc}
\end{figure*}

\begin{table*}
	\centering
		\begin{tabular}{l |c|c|c|c|c|c }
		\hline
  	\textbf{Model}& \textbf{$\rm n_{AM}~(cm^{-3})$}  & \textbf{\rm $m_{FS}$} & \textbf{CD:FS} & \textbf{RS:FS} & \textbf{Spectral Fitting ($\chi^2/dof$)} \\
  		\hline
  	 Observations &     &$0.59 \pm 0.03^a$ & $0.91 - 0.95^b$ &$0.68 - 0.75^b$ & \\
	 \hline
  	${\rm SNR_{ISM,dyn.,W7}}$       & 0.13 & 0.59 & 0.82 & 0.76 & 176 \\
         %\hline
         ${\rm SNR_{ISM,spectra,W7}}$        & 0.85 & 0.51 & 0.76 & 0.65 & 76 \\
         	 %\hline
        ${\rm SNR_{wind,W7}}$         & 0.13 & 0.61 & 0.82 & 0.77 & 140 \\
          \hline
  	${\rm SNR_{ISM,dyn,DDTc.}}$        & 0.13 & 0.58 & 0.83 & 0.76 & 172 \\
	%\hline
         ${\rm SNR_{ISM,spectra,DDTc}}$        & 1.4 & 0.50 & 0.76 & 0.63 & 53 \\
         %\hline
         ${\rm SNR_{wind,DDTc}}$        & 0.4 & 0.57 & 0.80 & 0.73 & 34 \\
     		\end{tabular}
		\caption{Comparison of the dynamical properties and the integrated spectra between the tree studied models and the current observations of Tycho. References: $^a$ \citet{katsuda10}; $^b$ \citet{warren05}. 
The observational data have been averaged for the studied region (see text for details) }
	\label{tab:3models}
\end{table*}

Fig. \ref{SN_comp} shows the density profile, the time evolution of the studied dynamical properties and the spectra of the three best models for the W7 explosion model. The SNR density profile and the spectra refer to the current age of Tycho ($t \sim 440$ yrs). 
Table~\ref{tab:3models} summarizes the values of the dynamical properties of each model at the current age of Tycho as well as the goodness-of-fit parameter ($\chi^2/dof$) for the comparison of the X-ray model spectra to
the observed spectra. 
Note that the $\chi^2/dof$ are from statistical point of view not acceptable,
but the spectra are also not the result of a normal fitting procedure.
Rather they serve as an indication on how well the observed spectra match 
the synthetic spectra.

The ${\rm SNR_{ISM,dyn}} $ model (left column) is set up to agree with the observations of the expansion parameter, and requires an ambient medium density of $n_{AM}= 0.13~\rm{cm^{-3}}$. 
Furthermore, the ratio RS:FS is close to the observed limits while this of CD:FS is rather low,
but see the discussion for potential caveats.
The resulting spectrum of this model deviates substantially from the observed one, being characterized by less strong lines  and different line centroids, as a result of the low ionization timescale of the shocked-ejecta shell.     
In contrast, the current X-ray spectrum of Tycho is best reproduced -on the base of the W7 explosion model-  by the simulation where the SNR is interacting with  a homogeneous ambient medium of $n_{AM}=0.85~\rm{cm^{-3}}$,  as shown in the second column of Fig~\ref{SN_comp}. However,  due to this high ambient medium density, the SNR has a lower expansion rate (0.51 instead of 0.59) and the ratios of CD:FS, RS:FS are too low, compared to the observations  (see central column of Fig. \ref{SN_comp} and Table \ref{tab:3models}).

\begin{figure}
	\centering
		\includegraphics[trim=20 10 20 10,clip=true,width=\columnwidth]{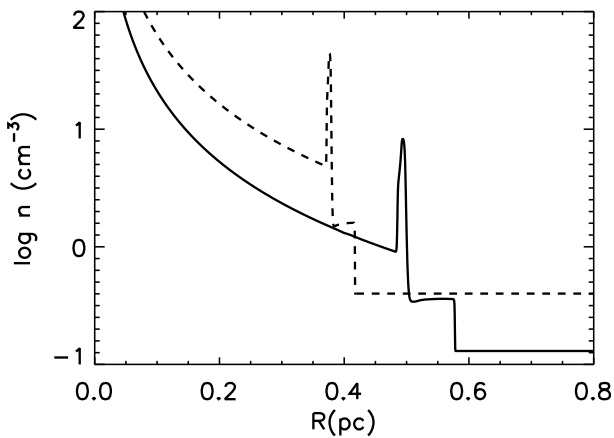}
	\caption{The density profile of the wind bubble at the moment of the SNR introduction that provides the best match with the observed properties of Tycho. The solid line corresponds to the case of the W7 explosion model and the dashed line to the DDTc.} 
	\label{fig:w7_bubble}
\end{figure}

The right column of Fig. \ref{SN_comp}  depicts the SNR model with a wind collision history that yields the best agreement with the studied observed properties of Tycho (${\rm SNR_{wind}} $ model). The wind bubble properties of this model are:  $\dot{M}=10^{-6}~ \rm{M_{\odot}~yr^{-1}}$,  $u_w= 10~ \rm {km s^{-1}}$ and $\tau_w= 5 \times 10^4$~yr. The ambient medium density that surrounds the bubble is $n_{AM} = 0.13~ \rm{cm^{-3}}$. The size of the bubble at the moment that we introduce the supernova ejecta is $r_w \sim 0.6$~pc and its density distribution at this moment is shown in Fig. \ref{fig:w7_bubble}.

The ${\rm SNR_{wind}} $ model  provides a good match with the dynamical properties of Tycho, at least as good as the ${\rm SNR_{ISM,dyn}}$ model. As  explained in Sec. \ref{sec:study} the collision history of the SNR with the small and dense wind bubble, leads to the formation of several pairs of reflected/transmitted shocks which result to  a non-monotonous time evolution of the expansion parameter (right column of Fig.  \ref{SN_comp}). However, when the radius of the SNR becomes few times bigger than the size of the wind bubble the dynamical properties of the forward shock tend to be similar with these of a SNR that was evolving in  a uniform ambient medium from the beginning. This can be seen in Table \ref{tab:3models} where at the age of Tycho, the dynamical properties of the ${\rm SNR_{wind}} $ and  ${\rm SNR_{ISM,dyn}}$  model   are almost identical. On the other hand,  this collision history forms a dense shocked-ejecta shell with peak densities and ionization ages higher than  the ${\rm SNR_{ISM,dyn}}$ model. Hence, the X-ray spectra of the  ${\rm SNR_{wind}} $ model is characterised by ionisation ages closer to those observed in Tycho.

Similar conclusions are drawn for the DDTc explosion model,  the results of which are shown in Fig~\ref{SN_compDDTc} and the parameters can be found alongside those of the W7 model in Table~\ref{tab:3models}. The best ${\rm SNR_{ISM,dyn}}$  model is derived for a (constant) ambient medium density of $n_{AM} = 0.13~ \rm{cm^{-3}}$ while the best  ${\rm SNR_{ISM,spectra}}$ for $n_{AM} = 1.4~ \rm{cm^{-3}}$ (see Fig. \ref{SN_compDDTc}). However, also in this case these two models (${\rm SNR_{ISM,dyn}}$, ${\rm SNR_{ISM,spectra}}$) are not able to reproduce the X-ray spectrum and the dynamics of Tycho, respectively (see Table \ref{tab:3models}). 

The wind properties of the  best  ${\rm SNR_{wind}} $ model are:  $\dot{M}=3 \times 10^{-6}~ \rm{M_{\odot}~yr^{-1}}$,  $u_w= 10~ \rm {km s^{-1}}$,
whereas the mass outflow phase lasted for  $\tau_w= 4 \times 10^4$~yr . The ambient medium density in which the wind bubble evolved was $0.4~\rm{cm^{-3}}$  and the resulting size of the the wind bubble is $\sim 0.4$~pc (Fig. \ref{fig:w7_bubble}). This model reproduces the observed dynamics and morphological and dynamical properties of Tycho, but also shows a good agreement between the observed and modeled X-ray spectrum.
The ambient medium density is in agreement with the upper limit of $n<0.6$~cm$^{-3}$ provided by \citet{cassam07}, based on the lack
of thermal X-ray emission from the shocked ambient medium shell. The density is somewhat larger
than the recent estimate by \citet{williams13} of $n\sim 0.1-0.2$~cm$^{-3}$ for the SWW region, which was based on modelling the infrared dust emission. Note,
however, that the infrared modelling contains some systematic uncertainties, and \citet{williams13} put more trust in the relative density contrast between
the western and eastern regions than in the absolute value of the density.

\subsection{Implications for distance estimate of Tycho's SNR}

Another problem for the uniform ambient medium model for Tycho's SNR is the distance estimates provided by the hydrodynamical/X-ray radiation
modeling.
 For a given model,  the distance of Tycho can be extracted either by comparing the radius and the velocity of the SNR's discontinuities (FS, CD, RS) with their observed angular radius and velocities (which we call geometrical distance),  or by equating the resulted total SNR flux of the model  with the observed flux of Tycho (photometrical distance). For a self-consistent model for Tycho,  these two methods of  distance calculations should converge. In addition, the distance estimate should be within the range of the other independent methods of Tycho's distance estimates ($1.8 -5$~ kpc ; see Introduction). 

For this purpose we calculate the distance of Tycho based on the observed angular radius of the forward, contact discontinuity and reverse shock (251'', 241'', 183'' respectively, Warren et al. 2005), the proper motion of the forward shock ($0.37~ \rm{arcsec~yr^{-1}}$, Katsuda et al 2010) and finally from the observed flux.  Note that the position and velocity of the forward shock can be influenced by the efficiency of cosmic-ray (CR) acceleration while the interface of the contact discontinuity is deformed by the Rayleigh- Taylor (R-T) instabilities that dominate in this region. These processes are not included in our 1D simulations, but will be discussed in %The importance  and the effects of these processes at the final outcome of the models will be discussed
Sect.~\ref{Sect:Discuss.}.

The estimated distances of Tycho, for the  three studied models are shown in Table \ref{tab:distances}. For both W7 and DDTc,  the ${\rm SNR_{ISM,dyn}}$ model, due to the chosen low ambient medium densities, forms an extended but non-dense SNR. Consequently, the geometrical distance is larger than the photometrical one especially for the DDTc model (1.1 - 1.6 times larger for the W7; 4.5 - 6.5 for the DDTc). The inverse is true for the ${\rm SNR_{ISM,spectra}}$ models. The small radius of the derived SNR coupled with the high density remnant's shocked shells lead to lower geometrical distances than the photometrical 
distance estimates (1.8 - 2.8 times lower for the W7; 1.3 - 1.9 for the DDTc). For the case of the ${\rm SNR_{wind}} $ models, a better convergence between the geometrical and photometrical distance is achieved where the photometrical distance for the W7 model is 1.03 - 1.4 higher than the geometrical estimation while for the DDTc is within the range of the geometrical distances.

\begin{table*}
	\centering
		\begin{tabular}{l |c|c|c|c|c|c|c }
		\hline
  	\textbf{Model}& \textbf{$ \rm n_{AM}~(cm^{-3})$}  & \textbf{ $ \rm D_{R_{FS}}$~(kpc)} & \textbf{ $ \rm D_{R_{CD}}$~(kpc)} & \textbf{$ \rm D_{R_{RS}}$~(kpc)} & \textbf{$\rm D_{V_{FS}}$~(kpc)} & \textbf{$\rm D_{flux}$~(kpc)} \\
           \hline
  	${\rm SNR_{ISM,dyn.,W7}}$        & 0.13 & 3.8 &3.3 & 4.0 & 2.8 & 2.5 \\ 
         % \hline
        ${\rm SNR_{ISM,spectra,W7}}$      & 0.85 & 2.9 & 2.3& 2.6 & 1.9 & 5.3 \\
       %  \hline
         ${\rm SNR_{wind,W7}}$       & 0.13 & 3.8 &3.3 & 4.0 & 2.9 & 4.1 \\
        \hline
  	${\rm SNR_{ISM,dyn.,DDTc}}$        & 0.13 & 3.7  &3.2  & 3.9 & 2.7 & 0.6 \\
      %  \hline
       ${\rm SNR_{ISM,spectra,DDTc}}$       & 1.4 & 2.8 & 2.1& 2.6 & 1.9 & 3.6 \\
     %   \hline
         ${\rm SNR_{wind,DDTc}}$         & 0.4 & 3.1 & 2.6 & 3.2 & 2.2 & 2.6 \\
     		\end{tabular}
	\caption{The extracted geometrical and emission brightness distance  of Tycho for the three studied models (see text for details).}
	\label{tab:distances}
\end{table*}

\section{Discussion}\label{Sect:Discuss.}

With our combined hydrodynamic/X-ray modeling of Tycho we test whether
an ambient medium structure more complex than a uniform density is able
to alleviate some of the discrepancies between Tycho's dynamical and X-ray
modeling. As we have indicated above it is indeed possible to obtain better
 and more self-consistent fits to the morphological, dynamical and observed X-ray
spectral properties of Tycho as compared to uniform AM models.

Our more complex model involved a circumstellar medium with a dense wind shell,
leading to an earlier development of the reverse shock, thereby
shocking the ejecta at an earlier time and with a higher density.
Although the match to the observed properties of Tycho is not perfect, this
more complex model has much less internal problems, in that
distance estimates based on X-ray flux and dynamics are consistent, and, 
moreover,the X-ray spectrum is well fitted.

However, our approach contained some simplifications, 
both in its initial assumption about the progenitor history and in the treatment
of the hydrodynamics.
To start with the latter; we assumed spherical symmetry, which is in itself
a reasonable assumption, given the  fact that we limited ourselves to the smooth
SWW part of the SNR. But by using 1D calculations we neglect the potential
effects of Rayleigh-Taylor instabilities. Rayleigh-Taylor instabilities
smooth out the high densities near the contact discontinuity, and mixes
shocked-ejecta material with shocked-circumstellar material;
see Fig. 3 and 4 of \citet{Wang2001} and Fig.4 of \citet{WarrenBlondin2013}.
As a result the shocked-ejecta shell is thicker up to $3 -10$ \%.
The dynamics and, more specifically, the expansion parameter of the forward 
shock is not affected by the presence of the Rayleigh-Taylor 
instabilities.
% unless the remnant is rather dynamically old and it reveals efficient CR acceleration.

Another assumption in our modeling is that efficient 
cosmic-ray acceleration does not significantly affect the dynamics, morphology 
and X-ray emission from the ejecta (a power law component
was added to all X-ray spectral models to account for synchrotron emission).
However, it is still an open question whether, and by how much, Tycho's dynamics and
morphology are affected
by cosmic-ray acceleration. Efficient cosmic-ray acceleration
changes the equation of state of the plasma, but also give rise to
energy flux losses due to escape of high energy cosmic rays, leading to
much higher compression ratios at the forward shock 
\citep[e.g.][]{decourchelle00,ellison04,Vink10a,kosenko11,WarrenBlondin2013}.
The proximity of ejecta material close to the forward shock appears
to support the idea of high compression factors \citep{warren05,Ferrard2010,WarrenBlondin2013}.

We did not include the effects of cosmic-ray 
acceleration, as we already have quite an extensive grid of models. However,
for the discussion on the circumstellar medium structure of Tycho
it is important to note that both efficient cosmic-ray acceleration and
Rayleigh-Taylor instabilities aggravates the problems of uniform ambient
medium models.  The reason is that in both cases the density of the
shocked-ejecta shell decreases. For Rayleigh-Taylor instabilities because
it partially removes the high density spike near the contact discontinuity.
For efficient acceleration, because the contact discontinuity
is brought close to the forward shock, and as a result the reverse shock is
at a larger radius, 
corresponding to a lower density at which ejecta are shocked.
A lower density of the shocked ejecta results in a lower ionization,
which reduces the consistency between the dynamics and the ionization age of Tycho.
Note that according to \citet{WarrenBlondin2013} the expansion parameter is only
mildly affected (at the 3\% level) 
by efficient cosmic-ray acceleration.\footnote{To see this
note that the Sedov self-similar solution, leading to $m=2/5$ for the Sedov 
case, uses only dimensional arguments, and not the equation of state of the 
plasma.The equation of state only enters as a linear numerical factor for
the shock radius as a function of $t$.
}

So to sum up, both Rayleigh-Taylor instabilities and efficient cosmic-ray 
acceleration may affect the dynamics and morphology of Tycho, but they 
aggravate
the problem of the low ionization age of the {\em ejecta}, requiring higher
ambient densities, which in turn creates a problem for the expansion 
parameter of Tycho. So they do not solve the problems we have tried to
address in this study.

\subsection{On the origin of the circumstellar medium around Tycho's SNR }

The CSM used in our models was formed by a slow ($u_w \sim 10~ \rm {km ~s^{-1}}$), short lived $\tau_w \sim 5 \times 10^4$~yrs wind with a mass loss rate of $\dot{M}\sim 10^{-6}~ \rm{M_{\odot}~yr^{-1}}$. 
An important question is to what type of Type Ia progenitor system this could be connected.

Circumstellar structures around SNe Ia progenitors are generally taken as evidence in favour of a SD scenario, as mass outflows are expected both from  the donor star and the accretor. However, the winds that emanate from the WD surface \citep{Hachisu1996,Hachisu1999} 
are inconsistent with our CSM models, as they are characterized by high terminal velocities of the order of  $ \sim 1000 ~ \rm {km s^{-1}}$. This kind of mass outflows forms  extended cavities instead of small, dense bubbles \citep{badenes07}. The same line of argument applies to main-sequence (MS) or subgiant donor stars as, due to their high gravitational potential, their outflows are in the form of fast, tenuous winds. Consequently, our models do not support  the identification of the subgiant star found at the centre of the historical  remnant \citep[Tycho G,][]{ruiz04} with the donor star in Tycho's progenitor system.  

The wind properties used in our models look rather similar to the stellar winds of low or intermediate mass giant stars.  However, if  Tycho's progenitor was a symbiotic binary, the surviving  donor star would be observable in the centre of the SNR, as giants are luminous stars. \citet{Kerzendorf2012} 
studied the properties of six stars that are lying in the geometrical center of Tycho, attempting to find distinct properties that could characterize a Type Ia companion star.   Among these six stars, there are two  giants with masses $2.4 \pm 0.8~ {\rm M_{\odot}}$ and $0.9 \pm 0.4~ {\rm M_{\odot}}$ (Tycho A and Tycho~C in their notation, respectively). Even if the distance of these stars are within the estimated limits of Tycho's distance, their low rotational velocity is in conflict with a Roche-lobe overflowing giant-donor-star scenario and only a wind accretion scenario is possible, as discussed by \citet{Kerzendorf2012}. 

 Another issue for our preferred model is that the mass outflow phase  ($\sim 5 \times 10^4$~ yrs) is about one or two orders of magnitude smaller than the life time of a giant ($\tau_{giant} \sim 10^5 - 10^6$~yrs, depending on the initial mass and metallicity). This requires
that for our model the WD exploded 
almost right after ($\Delta t \sim 5 \times10^4$ yrs) the evolution of the donor star to the giant branch. Although this is a possibility, it requires fine tuning,
and in addition it requires that an efficient mass transfer process took place at a previous stage of the binary evolution, or a sub-Chandrasekhar
explosion.

Other possible mass outflows  scenarios may not require winds from the donor star. For example, 
 nova explosions have been shown to accompany the final phase of the binary evolution in several SNe Ia progenitor systems
 \citep[see e.g.][]{Hachisu2008}. The low ejecta mass ($M_{ej} \sim 10^{-7} ~{\rm M_{\odot}}$) and high velocity (${\rm few \times 10^3~ km~s^{-1}}$) of a nova explosion form a small cavity around the system, something that contradicts with the results of our models. But a sequence of nova explosions,
 in which the nova shells collide and interact with each other, could possibly form an ambient medium with properties similar to that of the CSM used in our models. Further investigation and simulations of such a system are necessary in order to draw a firm conclusion on whether this can
 provide similar observational characteristics as the small, dense wind bubble that we used in this paper.

Finally, there is the possibility that the CSM that surrounds Tycho has a DD origin. Population synthesis models of DD SNe Ia have shown  that about $0.5 \%$ of these systems explode during or right after ($\Delta t < 1000 $~yrs) the last common envelope (CE) episode \citep{Ruiter2013}. A similar conclusion is also extracted from the core-degenerate scenario of SNe Ia \citep{Ilkov2012,Soker2011}. Could this (semi-) ejected common envelope reveal similar properties with the CSM used in our model? 
The mass ejection process during the CE phase is still not clear, and has been difficult to model,
but 
recent simulations show that the expected mass loss rate during the common envelope is of the order of ${M}\sim 10^{-3} - 10^{0} ~ \rm{M_{\odot}~yr^{-1}}$ and the phase of this mass outflow is varying from a few months to years \citep{Ricker2012}. Thus, in this scenario a denser and smaller CSM is expected compared to the CSM used in our models.

\section{Conclusions}

We have modelled the structure, evolution and X-ray emission of a Type Ia SNR resembling Tycho's SNR. 
 We have shown that a delayed detonation explosion model combined with a more
complex CSM, namely a CSM modified by the wind of the progenitor system,  can solve some
of the discrepancies in the models of \citet{badenes06}. These discrepancies are that the high ionization age of the shocked-ejecta requires
a high ambient medium density ($n \sim 0.9$~cm$^{-3}$), which is at odds with measured dynamics of the SNR, which requires 
$n \lesssim 0.2$~cm$^{-3}$ \citep{reynoso97,katsuda10}.

The reason that a wind bubble model solves the discrepancy is that a relative dense wind bubble in the central parsec surrounding
the explosion, triggers an early formation of the reverse shock, thereby increasing the ionisation age of the ejecta. On the other
hand, the compactness of the wind bubble ensures that currently the shock wave is moving through less dense, undisturbed, ambient medium,
thereby satisfying the measured expansion parameters.
The best-matching wind bubble model has  a progenitor 
mass loss rate of $\dot{M}\sim 10^{-6}$~M$_{\odot}$/yr and wind velocity of $u_{\rm w}=10$~km/s. The
mass-loss phase should be of the order of $10^4-10^5$ yr, as then the shock wave of Tycho must have passed beyond the
wind bubble and is moving through the ambient medium.

For our models we used the ejecta structure and kinematics of the both the W7 deflagration model \citep{nomoto84}
and the DDTc delayed detonation model \citep{bravo2005,badenes05}. The DDTc wind model 
gave the best fit  to the X-ray spectra of Tycho while it reproduces its observed expansion parameter. The density of the ISM
used in this model is 0.4~cm$^{-3}$ which is within the limit placed by the lack of thermal X-ray emission behind Tycho's forward shock \citep{cassam07},
but 2-4 times higher than estimates based on modeling of infrared 
observations of the remnant\citep{williams13}.  
The model based on the DDTc explosion also provides a distance estimate of
$2.7\pm 0.5$~kpc, which is both
estimated from the total X-ray luminosity and from comparing the predicted radius and velocity 
to the angular radius of the SNR. This estimate is in agreement with independent observational estimates.

Note that apart from the match between the synthetic spectrum and the observed
spectrum, the W7 model with a bubble also provide good results. However,
the W7 models do not perform well when it comes to the spectral characteristics.
An obvious discrepancy between the model and the data is that the W7
model predicts brighter magnesium (Mg XI K$\alpha$) emission than observed
\citep[see also the model fits in][]{badenes06}.

We contrast our best-matching model with best-matching uniform ISM  models  similar to the one used by \citet{badenes06}, also both
with the  W7 and DDTc Type Ia models. These models confirm the discrepancies in ambient medium densities
between models that best fit the X-ray spectra ($n \sim 0.8 -1.4$~cm$^{-3}$) and models that best fit the measured expansion parameter
($n \sim 0.13$~cm$^{-3}$).

 Although the DDTc model with wind-loss from the progenitor system does give a better description of the observed properties of Tycho,
our model is likely a simplification of the Tycho SNR.
First of all we neglected the variation in shock and emission properties around the SNR. We circumvented this variation by
concentrating on the more uniformly appearing southwest-west region of the remnant, following also the work of \citet{badenes06}.
In addition, our  model could  be further extended by including the potential effects of cosmic-ray acceleration, hydrodynamic
instabilities, and potential density gradients, the latter two complications requiring 2D or 3D simulations. However, we have argued that including
cosmic-ray acceleration and hydrodynamic instabilities  does not solve the discrepancies between ionisation age and
shock kinematics that characterise SNR models involving a uniform ambient medium.

Another open question related to our best-fit CSM model is that the wind-loss history of the progenitor
is difficult to connect
to existing realistic stellar evolution models of the binary system, as the wind-loss phase is much shorter than may be expected from a giant donor
star. 
In this context it is worthwhile investigating whether a similar CSM structure can be obtained by a sequence of nova explosions prior
to the Type Ia explosion, creating a cavity surrounded by a denser shell, which may be expected in the single degenerate scenario.
Or alternatively, whether a CSM altered by mass loss during a common envelope phase under the assumption of double degenerate
scenario, could be responsible for a higher density close the explosion centre.
\\
\\
\textit{Acknowledgments.}  
We are grateful to Carles Badenes and Eduardo Bravo for providing us with the DDTc Type Ia model, Rony Keppens for his help
with the AMRVAC code, and Sergei Blinnikov for the use of the SUPREMA code. We also thank Anne Decourchelle  for carefully reading a draft version of this paper and for her helpful suggestions that have helped us improve the manuscript. DK contribution was supported by Dutch Óopen competitionÓ grant and by the French national research agency (ANR) grant COSMIS. KMS has received funding
from grant number ST/H001948/1
made by the UK Science Technology and Facilities Council.

%\bibliographystyle{mn2e}
%\bibliography{sn,snrs,article1}

\label{lastpage}
\end {document}